\documentclass[linenumbers]{aastex61}
\usepackage{lineno}
\nolinenumbers
\received{}
\revised{}
\accepted{\today}
\submitjournal{ApJ}

\shorttitle{Planet Formation at Pressure Bumps}
\shortauthors{Chambers}

\newcommand{\ain}{a_{\rm in}}
\newcommand{\aout}{a_{\rm out}}
\newcommand{\fgap}{F_{\rm gap}}
\newcommand{\hgas}{H_{\rm gas}}
\newcommand{\hpeb}{H_{\rm peb}}
\newcommand{\lpeb}{L_{\rm peb}}
\newcommand{\matmos}{M_{\rm atmos}}

\newcommand{\mstar}{M_\ast}
\newcommand{\period}{P_{\rm orb}}
\newcommand{\rcap}{r_{\rm cap}}
\newcommand{\rset}{r_{\rm set}}
\newcommand{\sigmagas}{\Sigma_{\rm gas}}
\newcommand{\sigmanogap}{\Sigma_{\rm no\ gap}}
\newcommand{\sigmapeb}{\Sigma_{\rm peb}}
\newcommand{\st}{\rm St}
\newcommand{\stcrit}{{\rm St}_{\rm crit}}
\newcommand{\tdrag}{t_{\rm drag}}
\newcommand{\tgas}{t_{\rm gas}}

\newcommand{\twave}{t_{\rm wave}}
\newcommand{\vfrag}{v_{\rm frag}}
\newcommand{\vesc}{v_{\rm esc}}
\newcommand{\vgas}{v_{\rm gas}}
\newcommand{\vkep}{v_{\rm kep}}
\newcommand{\vrel}{v_{\rm rel}}
\newcommand{\wgap}{w_{\rm gap}}

\begin{document}

\title{Rapid Formation of Jupiter and Wide-Orbit Exoplanets in Disks with Pressure Bumps}

\correspondingauthor{John Chambers}
\email{jchambers@carnegiescience.edu}

\author{John Chambers}
\affil{Carnegie Institution for Science \\
Earth and Planets Laboratory, \\
5241 Broad Branch Road, NW, \\
Washington, DC 20015, USA}

%
%
\begin{abstract}
\nolinenumbers
The formation of gas-giant planets within the lifetime of a protoplanetary disk is challenging especially far from a star. A promising model for the rapid formation of giant-planet cores is pebble accretion in which gas drag during encounters leads to high accretion rates. Most models of pebble accretion consider disks with a monotonic, radial pressure profile. This causes a continuous inward flux of pebbles and inefficient growth. Here we examine planet formation in a disk with multiple, intrinsic pressure bumps. In the outer disk, pebbles become trapped near these bumps allowing rapid growth under suitable conditions. In the inner disk, pebble traps may not exist because the inward gas advection velocity is too high. Pebbles here are rapidly removed. In the outer disk, growth is very sensitive to the initial planet mass and the strength of turbulence. This is because turbulent density fluctuations raise planetary eccentricities, increasing the planet-pebble relative velocity. Planetary seeds above a distance-dependent critical mass grow to a Jupiter mass in 0.5--3 million years out to at least 60 AU in a 0.03 solar-mass disk. Smaller bodies remain near their initial mass, leading to a sharp dichotomy in growth outcomes. For turbulent alpha = 1e-4, the critical masses are 1e-4 and 1e-3 Earth masses at 9 and 75 AU, respectively. Pressure bumps in disks may explain the large mass difference between the giant planets and Kuiper belt objects, and also the existence of wide-orbit planets in some systems.

\end{abstract}

%
%
\section{Introduction}

Over the last two decades, thousands of extrasolar planetary systems have been discovered \citep{cumming:2008, foreman-mackey:2016, petigura:2018, wittenmyer:2020}. As a result, much recent work on planet formation has focussed on understanding the origin and characteristics of these systems. However, it is worth emphasizing that we still lack a satisfactory model for the formation of our own planetary system even though we know much more about the Sun's planets than those in other systems, and the Solar System has been studied for far longer.

The Solar System may be something of an outlier in the distribution of planetary systems. Among its more obvious features are the presence of multiple giant planets orbiting far from the Sun, a lack of planets larger than Earth inside 5 AU, and a complete lack of planets interior to 0.3 AU. While the frequency of distant giant planets in other systems is still uncertain, it is clear that super-Earths in short-period orbits are common \citep{burke:2015, petigura:2018}. This makes the absence of similar planets in the Solar System noteworthy.

A particular sticking point in understanding the formation of the Solar System is the origin of the giant planets. Early work on this problem looked at the growth of the giants' solid cores by runaway and oligarchic growth within a population of planetesimals \citep{lissauer:1987, inaba:2003, thommes:2003, ida:2004}. However, recent studies suggest that cores massive enough to accrete large gaseous envelopes are unlikely to form at the orbital distances of the giant planets within the typical lifespan of a protoplanetary disk \citep{levison:2010, fortier:2013, johansen:2019}. This difficulty is even more acute when it comes to forming the population of extrasolar giant planets with wide orbits \citep{dodson-robinson:2009} found by direct imaging \citep{marois:2008, haffert:2019}.

A plausible alternative for giant-planet formation is the pebble accretion model. Here, planetary embryos grow rapidly by sweeping up mm-to-dm sized ``pebbles'', aided by gas drag during an encounter between a pebble and the embryo \citep{ormel:2010, lambrechts:2012}. Unfortunately, while pebble accretion may explain the origin of giant planets, it does not explain why the Sun's outer planets grew very massive while the inner ones remained small.

To explain this dichotomy, \citet{morbidelli:2015} proposed that planetary growth rates inside and outside the ice line of the solar nebula were different because the pebbles in the two regions were different. Pebbles outside the ice line may have grown larger due to the different sticking properties of ices versus silicates \citep{bridges:1996}. The large, icy pebbles were swept up rapidly leading to the formation of large giant-planet cores beyond the ice line. In contrast, pebbles inside the ice line remained small, and hard to capture, retarding the growth of planets in this region.

This idea seems promising, but there are two potential problems. Firstly, it doesn't explain why the same processes led to a very different outcome in many other planetary systems.  Secondly, although large planets may begin forming just beyond the ice line, they are likely to migrate inwards due to tidal interactions with the gas disk. If the ice line was a few AU from the Sun \citep{lecar:2006, kennedy:2008}, large planets could easily migrate into the terrestrial-planet region, leading to the presence of super Earths or giant planets where we don't see them. Simulations of planet formation with migration suggest that in order to form giants with orbits like Jupiter, the cores of these planets had to form several tens of AU from the Sun \citep{coleman:2014, bitsch:2015, johansen:2019b}. Significant migration is likely even when pebble accretion allows rapid growth. Large initial orbits give planets room to migrate inwards as they grow, so that they are still relatively far from the star when migration ceases.

Pebbles in protoplanetary disks tend to undergo rapid radial drift, moving towards high pressure regions \citep{weidenschilling:1977a}. In a monotonically varying disk this means pebbles drift inwards. In the Solar System, pebble accretion could have formed large outer planets and small inner ones if the flux of pebbles into the inner disk was terminated at an early stage. The existence of a pebble barrier would naturally explain why most meteorite parent bodies appear to belong to two distinct populations with different isotopic compositions \citep{kruijer:2017}. This difference could have arisen if the two populations formed mainly from material on either side of the barrier.

\citet{lambrechts:2014a} have noted that the formation of Jupiter's core could have stopped the inward drift of pebbles into the inner solar system by generating a pressure maximum in the disk just outside the planet's orbit. This would naturally stunt the growth of inner planets by terminating pebble accretion, leaving only the slower planetesimal growth mode. However, there is timescale problem. Models suggest that the formation of Jupiter's core, in a monotonically varying disk, would take too long to be compatible with the onset of the meteorite dichotomy determined by radiogenic dating \citep{brasser:2020}. 

The meteorite dichotomy could still be due to a localized pressure maximum if this bump were an intrinsic feature of the disk that arose at an early stage, rather than caused by a planet \citep{brasser:2020}. The existence of this pressure bump would retard the growth of planets interior to it while allowing giant planets to form further out.

ALMA observations show that many protoplanetary disks have non-monotonic pebble distributions, with rings or spiral structures being very common \citep{andrews:2018}. It is plausible that these pebble structures are themselves caused by localized pressure bumps. A relatively modest pressure bump can stop the inward drift of pebbles leading to a ring as incoming pebbles pile up at the bump \citep{haghighipour:2003, kretke:2007, dullemond:2018}.
In fact, models of pebble orbital and collisional evolution in disks with long-lived pressure bumps predicted the existence of ring-like features before they were detected by ALMA \citep{pinilla:2012}.

It is still unclear whether the bumps seen in disks are typically caused by planets that have already formed or whether they are intrinsic properties of the disk. In the latter case, it is not known how long the pressure features last. What is clear is that a large fraction of disks show these features and they should be considered in models for planet formation.

\citet{morbidelli:2020} has examined the growth of protoplanets at one of the pressure bumps observed by ALMA. This bump is located roughly $77\pm 5$ AU from the star Elias 24. He found that growth by pebble accretion occurs, but it rather slow in this case.

In this paper, we extend this scenario to look at planet formation in disks with multiple bumps spanning the entire disk. We take no position on the mechanism that generates pressure bumps, merely noting that they are a common feature of real disks, and examining their effect on planet formation. 

We find that giant planet formation can be very effective, forming cores large enough to accrete gas within a few hundred thousand years. In our model, giant planets tend to form at intermediate distances in the disk. Growth is very slow in the inner and outer disk, for different reasons that we describe in detail. This naturally gives rise to systems like the Solar System with giant planets sandwiched between small terrestrial planets and a Kuiper belt of residual protoplanets.

However, the outcome in the outer disk is very sensitive to the the initial protoplanet mass and the turbulence level, as well as the height of the pressure bumps. For example, increasing the initial protoplanet mass by an order of magnitude above our baseline case (from Ceres to Pluto mass) readily allows gas-giant planets  to form out to at least 60 AU, comparable to the orbital distances of some directly imaged planets \citep{marois:2008, haffert:2019}. This sensitivity suggests that a wide diversity of planetary systems could arise due to relatively small differences in the conditions from one disk to another.

The rest of this paper is organized as follows. In Section~2, we describe the model used to calculate the growth and orbital evolution of the planets. The results are described in Section~3. The implications and caveats are discussed in Section~4. Finally, Section~5 contains a summary.

%
%
\section{The Model}
In this section, we describe the model for planet formation used in this study.

%
%
\subsection{Gas Disk}
We consider the growth and orbital evolution of a small number of planetary embryos in an evolving protoplanetary disk with a series of stationary bumps. The gas surface density (in the absence of planetary gap opening) and temperature at a distance $a$ from the star are given by
\begin{eqnarray}
\sigmanogap&=&\Sigma_0\left(\frac{a}{a_0}\right)^{-1}
F(a)\exp\left(-\frac{t}{\tgas}\right)
\hspace{15mm} \ain<a<\aout
\nonumber \\
T&=&T_0\left(\frac{a}{a_0}\right)^{-1/2}
\label{eq-sigmagas}
\end{eqnarray}
where $a_0=1$ AU. Here, $\ain$ and $\aout$ are fixed, and $\Sigma_0$, $T_0$ and $\tgas$ are model parameters. The function $F(a)$ describes the bump profile. Note that we have deliberately chosen a simple underlying disk model in order to make the effect of the bumps clearer.

The surface density normalization factor $\Sigma_0$ is related to the initial disk mass $M_{\rm disk,0}$ by
\begin{equation}
M_{\rm disk,0}\simeq 2\pi\Sigma_0a_0\aout
\end{equation}

The bumps are modeled using a sinusoidal function with the bumps logarithmically spaced in distance
\begin{equation}
F(a)=1+B\sin\left[\omega\ln\left(\frac{a}{\ain}\right)-\pi
\right]
\label{eq-fbump}
\end{equation}
where $B$ and $\omega$ are model parameters, and the phase of the sine function is chosen to ensure that the innermost maximum does not coincide with the inner edge of the disk.

%
%
\subsection{Pebbles}
The disk contains pebbles that can be accreted by the embryos. Following \citet{dullemond:2018}, pebbles drift radially with a velocity $v_r$ given by
\begin{equation}
v_r=
\frac{\vkep\st}{(1+\st^2)}
\left(\frac{c_s}{\vkep}\right)^2\frac{d\ln P}{d\ln a}
+\frac{\vgas}{(1+\st^2)}
\label{eq-vr}
\end{equation}
where $\vkep$ and $c_s$ are the Keplerian velocity and sound speed in the gas, $P$ is the gas pressure, and $\st$ is the Stokes number of the pebbles. The first term on the righthand side of this equation represents the drift of the pebbles towards the nearest pressure maximum due to a headwind or tailwind caused by a radial pressure gradient. The second term represents the inward advection of pebbles driven by the flow of gas towards the star with a velocity $\vgas$, given by
\begin{equation}
\vgas=-\frac{1}{2\pi a\sigmanogap}\frac{M_{\rm disk}}{\tgas}
=-\frac{1}{F(a)}\frac{\aout}{\tgas}
\end{equation}
We examine the effect of neglecing the $\vgas$ term in Eqn.~\ref{eq-vr} in Section~3.7.

Pebbles also diffuse radially due to turbulent motions in the gas. Following \citet{dullemond:2018}, we assume that the diffusion coefficient is given by
\begin{equation}
D=\frac{\alpha_v c_s\hgas}{1+\st^2}
\end{equation}
where $\hgas$ is the gas scale height, and $\alpha_v$ is a model parameter describing the strength of the turbulence. Note that we do not assume that turbulence is the only factor driving disk evolution, so $\alpha_v$ and $\tgas$ are not necessarily related to each other.

The pebble Stokes number is determined by assuming that pebbles grow by sweeping up dust and other pebbles until further growth is halted by destructive collisions caused by turbulent motions. Following \citet{lambrechts:2014b}, this implies that
\begin{equation}
\st=\frac{1}{3\alpha_v}\left(\frac{\vfrag}{c_s}\right)^2
\label{eq-st}
\end{equation}
where $\vfrag$ is a model parameter describing the collision speed at which pebbles begin to fragment. We use $\vfrag=100$ cm/s, which is a typical value found in laboratory experiments \citep{guttler:2010}.

We calculate the advection and diffusion of the pebbles numerically using a radial grid with 1024 logarithmically spaced zones. The magnitude and sign of the pebble advection velocity depend on the local pressure gradient which is affected by the disk bumps and also by partial gaps in the gas disk opened by large embryos (described below).

The disk begins with no pebbles. At each disk location, pebbles are added at a single epoch designed to simulate the time taken for pebbles to grow from the small dust grains that are initially present. The pebble surface density at this epoch is set to $\sigmapeb=Z\sigmagas$, where $Z$ is the initial dust-to-gas ratio. Following \citet{lambrechts:2014b}, we assume that the pebble growth rate is controlled by turbulent motions. This leads to a pebble formation time $t_{\rm peb}$ given by
\begin{equation}
t_{\rm peb}\propto\frac{\period}{Z}
\end{equation}
where $\period$ is the orbital period. We choose the constant of proportionality so that pebbles are added after 400 orbital periods in our model.

%
%
\subsection{Pebble Accretion}
Following \citet{ormel:2010}, an embryo accretes pebbles so that its mass $M$ increases as
\begin{equation}
\left(\frac{dM}{dt}\right)_{\rm peb}=
\sigmapeb\vrel\times
\min\left[2\rcap,\frac{\pi\rcap^2}{2\hpeb}
\right]
\label{eq-dmdt-peb}
\end{equation}
where $\hpeb$ is the pebble scale height, set by turbulent diffusion and given by
\begin{equation}
\hpeb=\hgas\left(\frac{\alpha_v}{\alpha_v+\st}\right)^{1/2}
\end{equation}
\citep{youdin:2007}.

We use the following expression for the relative velocity $\vrel$ between a pebble and an embryo
\begin{equation}
\vrel=\max[v_{\rm OK}, e\vkep, i\vkep]
\label{eq-vrel}
\end{equation}
where $v_{\rm OK}$ is the relative velocity described by \citet{ormel:2010}, appropriate for an embryo moving on a circular orbit, and $e$ and $i$ are the orbital eccentricity and inclination of the embryo.

The capture radius, $\rcap$ is given by
\begin{equation}
\rcap=\rset\exp\left[-\left(\frac{\st}{\stcrit}\right)^{0.65}\right]
\label{eq-rcap}
\end{equation}
where $\rset$ is the capture radius for a pebble undergoing strong gas drag (the ``settling regime''), given by expressions in \citet{ormel:2010}. The exponential factor in this equation allows for the fact that pebbles with Stokes numbers exceeding a critical value are only partially affected by gas drag and are less likely to be captured than those in the settling regime. The critical Stokes number is given by
\begin{equation}
\stcrit=\min\left[1,
4\left(\frac{M}{\mstar}\right)
\left(\frac{\vkep}{\vrel}\right)^3\right]
\label{eq-stcrit}
\end{equation}

The expressions for $\vrel$ and $\rcap$ from \citet{ormel:2010} are given in Appendix~A.

When $\st>\stcrit$, gas drag is weak and we also have to consider whether pebble accretion by gravitational focussing is more important than settling. The gravitational capture radius is given by:
\begin{equation}
r_{\rm grav}=r_c\left[1+\left(\frac{\vesc}{\vrel}\right)^2\right]^{1/2}
\label{eq-rgrav}
\end{equation}
where $r_c$ is the physical radius, $\vesc$ is the escape velocity, and $\vrel$ is the relative velocity given by
\begin{equation}
\vrel=\max\left[v_{\rm OK},e\vkep,i\vkep,
\frac{r_H\vkep}{a}\right]
\end{equation}
and $r_H$ is the Hill radius of the embryo

In general, the capture radius is
\begin{eqnarray}
r&=&\rcap \hspace{37mm} \st<\stcrit
\nonumber \\
&=&\max[\rcap,r_{\rm grav}] 
\hspace{20mm} \st>\stcrit
\end{eqnarray}

%
%
\subsection{Gas Accretion}
Embryos accrete gas in one of two regimes. For low-mass embryos, the gas accretion rate is controlled by the rate at which the embryo's envelope can cool. We calculate the gas accretion rate in this case following \citet{bitsch:2015}
\begin{equation}
\left(\frac{dM}{dt}\right)_{\rm cool}=4.37\times 10^{-9}
\left(\frac{\kappa}{{\rm cm}^2/g}\right)^{-1}
\left(\frac{\rho_c}{5.5\ g/{\rm cm}^3}\right)^{-1/6}
\left(\frac{M_c}{M_\oplus}\right)^{11/3}
\left(\frac{M_e}{M_\oplus}\right)^{-1}
\left(\frac{T}{\rm 81\ K}\right)^{-1/2}
M_\oplus/{\rm y}
\label{eq-dmdtcool}
\end{equation}
where $M_c$ and $\rho_c$ are the mass and density of the embryo's core, $M_e$ and $\kappa$ are the mass and opacity of the envelope, and $T$ is the local disk temperature.

The gas accretion rate is reduced or stopped if the embryo is also accreting pebbles because both processes can balance the heat radiated away by the envelope. We describe how the gas accretion rate is modified by pebble accretion in Appendix B.

For massive embryos, the gas accretion is limited by the rate at which gas flows towards the embryo. To calculate this hydrodynamic accretion rate, we follow \citet{tanigawa:2016}
\begin{equation}
\left(\frac{dM}{dt}\right)_{\rm hydro}=
0.29\sigmagas a_p\vkep
\left(\frac{M}{\mstar}\right)^{4/3}
\left(\frac{a_p}{\hgas}\right)^2
\end{equation}
where $M$ and $a_p$ are the mass and semi-major axis of the planet, and $\sigmagas$ is the gas surface density including the effects of a partial gap cleared by the embryo.

The actual gas accretion rate is calculated using the lesser of the two accretion rates given above.

%
%
\subsection{Gap Opening}
Massive embryos open a partial gap in the gas disk. To calculate the depth $\fgap$ and width $\wgap$ of the gap, we follow \citet{kanagawa:2018}
\begin{eqnarray}
\fgap&=&\frac{1}{1+0.04K}
\nonumber \\
\wgap&=&\frac{a_p}{4}
\left(\frac{M}{\mstar}\right)^{1/2}
\left(\frac{a_p}{\hgas}\right)^{3/4}
\left(\frac{1}{\alpha_v}\right)^{1/4}
\end{eqnarray}
where
\begin{equation}
K=\left(\frac{M}{\mstar}\right)^2
\left(\frac{a_p}{\hgas}\right)^5
\frac{1}{\alpha_v}
\label{eq-bigk}
\end{equation}

We use the following expression for the radial profile of the gap
\begin{equation}
\frac{\sigmagas}{\sigmanogap}=
1-(1-\fgap)\exp\left[
-\frac{1}{4}\left(\frac{a-a_p}{\wgap}\right)^4\right]
\end{equation}
where $\sigmanogap$ is given by Eqn.~\ref{eq-sigmagas}. This profile approximately mimics the gap shapes calculated by \citet{duffell:2020}.

%
%
\subsection{Embryo Orbital Evolution}
The orbital evolution of the planetary embryos is calculated using the {\it Mercury\/} $N$-body integrator \citep{chambers:1999}, modified to include forces due to gas drag, tidal torques from the gas disk, and gravitational perturbations due to turbulent density fluctuations. All integrations use a stepsize of 5 days.

Changes in the orbital semi-major axis $a$, eccentricity $e$, and inclination $i$ due to interactions with the gas are modeled by applying the following accelerations to each embryo
\begin{equation}
\frac{d{\bf v}}{dt}=
\left(\frac{1}{2a}\frac{da}{dt}\right){\bf v}
+\left(\frac{1}{e^2}\frac{de^2}{dt}\right)
\frac{({\bf r}\cdot{\bf v})}{r^2}{\bf r}
+\left(\frac{1}{i^2}\frac{di^2}{dt}\right)
(0,0,v_z)
\end{equation}
where ${\bf r}$ and ${\bf v}$ are the coordinate and velocity vectors of the embryo.

Following \citet{adachi:1976}, the changes due to gas drag are
\begin{eqnarray}
\frac{de^2}{dt}&=&-\frac{2e^2}{\tdrag}(\eta^2+e^2+i^2)^{1/2}
\nonumber \\
\frac{di^2}{dt}&=&-\frac{i^2}{\tdrag}(\eta^2+e^2+i^2)^{1/2}
\end{eqnarray}
where
\begin{equation}
\eta=\frac{\vgas-\vkep}{\vkep}
\end{equation}
where $\vgas$ is the azimuthal gas velocity, and
\begin{equation}
\tdrag=\frac{6\rho r}{\sigmagas\vkep}
\end{equation}
where $r$ and $\rho$ are the radius and density of the embryo.

Following \citet{kobayashi:2018}, the changes due to turbulent density fluctuations in the gas are
\begin{eqnarray}
\frac{de^2}{dt}&=&0.0311\alpha_v
\left(\frac{\sigmagas a^2}{\mstar}\right)^2\frac{\vkep}{a}
\nonumber \\
\frac{di^2}{dt}&=&10^{-4}\frac{de^2}{dt}
\label{eq-turb-excite}
\end{eqnarray}
Note that these fluctuations excite $e$ and $i$ rather than causing damping.

Following \citet{ida:2020}, the changes in $e$ and $i$ due to tidal torques are
\begin{eqnarray}
\frac{de^2}{dt}&=&-\frac{0.780e^2}{\twave}
\left[1+\frac{1}{15}(e^2+i^2)^{3/2}\right]^{-1}
\nonumber \\
\frac{di^2}{dt}&=&-\frac{0.544i^2}{\twave}
\left[1+\frac{2}{43}(e^2+i^2)^{3/2}\right]^{-1}
\end{eqnarray}
where
\begin{equation}
\twave=\left(\frac{\mstar}{M}\right)
\left(\frac{\mstar}{\sigmagas a^2}\right)
\left(\frac{\hgas}{a}\right)^4
\frac{a}{\vkep}
\end{equation}

To calculate orbital migration due to tidal torques, we follow \citet{paardekooper:2010} and \citet{kanagawa:2018}. We assume that corotation torques are fully unsaturated, and examine the effect of relaxing this assumption in Section~3.8.

The change in semi-major axis is given by
\begin{equation}
\frac{da}{dt}=\left(\frac{da}{dt}\right)_L
+\left(\frac{da}{dt}\right)_C\exp\left(-\frac{K}{20}\right)
\label{eq-dadt}
\end{equation}
where $K$ is given by Eqn.~\ref{eq-bigk}. The first and second terms on the righthand side of the equation represent terms due to Lindblad resonances and corotation terms, respectively, given by
\begin{eqnarray}
\left(\frac{da}{dt}\right)_L&=&
(-2.5+0.1\phi-1.7\beta)\left(\frac{da}{dt}\right)_0
\nonumber \\
\left(\frac{da}{dt}\right)_C&=&
(1.65-1.1\phi+0.8\beta)\left(\frac{da}{dt}\right)_0
\label{eq-dadt-lc}
\end{eqnarray}
where
\begin{eqnarray}
\phi&=&-\frac{d\ln\sigmanogap}{d\ln a}
\nonumber \\
\beta&=&-\frac{d\ln T}{d\ln a}=\frac{1}{2}
\label{eq-phi-beta}
\end{eqnarray}
and
\begin{equation}
\left(\frac{da}{dt}\right)_0=
2\left(\frac{M}{\mstar}\right)
\left(\frac{\sigmagas a^2}{\mstar}\right)
\left(\frac{a}{\hgas}\right)^2\vkep
\label{eq-dadt0}
\end{equation}
Note that this formula uses the gas surface density $\sigmagas$ that includes the effect of partial gap opening.

Collisions between embryos are treated as mergers, conserving mass and linear momentum.

\startlongtable
\begin{deluxetable}{lll}
\tablecaption{Baseline Model Parameters}
\tablehead{
\colhead{Parameter} & \colhead{Symbol} & \colhead{Value} \\
}
\startdata
Stellar mass & $\mstar$ & $1 M_\odot$ \\
Initial gas disk mass & $M_{\rm disk,0}$ & $0.03M_\odot$ \\
Disk inner edge & $\ain$ & 0.4 AU \\
Disk outer edge & $\aout$ & 100 AU \\
Gas decay timescale & $\tgas$ & 1 My \\
Temperature at 1 AU & $T_0$ & 200 K \\
Turbulence parameter & $\alpha_v$ & $10^{-4}$ \\
Bump height & $B$ & 0.47 \\
Bump frequency & $\omega$ & $2\pi/\ln 2$ \\
Pebble fragmentation velocity & $\vfrag$ & 100 cm/s \\
Initial embryo mass & $M_0$ & $2\times 10^{-4}M_\oplus$ \\
Embryo bulk density & $\rho_c$ & 3 g/cm$^3$ \\
Envelope opacity & $\kappa$ & 0.1 cm$^2$/g \\
Initial rock/gas ratio & & 0.005 \\
Initial ice/rock ratio & & 1 \\
$N$-body integrator timestep &  & 5 days \\
Simulation duration & & 3 My \\
\enddata
\end{deluxetable}

%
%
\subsection{Initial Conditions}
We consider the formation of planets in a 0.03 solar mass protoplanetary disk surrounding a solar mass star. The inner and outer edges of the disk are fixed at 0.4 and 100 AU, respectively. Simulations last for 3 million years. The parameters describing the structure and evolution of the disk and other aspects of the model are listed in Table~1. In the next section, some of these parameters will be varied. In these cases, the values given in Table~1 apply to the baseline model.

The gas disk has 8 logarithmically spaced bumps. A single planetary embryo begins in each of these bumps, with the exact location corresponding to a stable equilibrium point for pebble drift, if it exists, or at the point where the magnitude of the radial drift is minimized. Since pebbles will tend to accumulate at these locations, we assume that embryos will naturally form here too. The initial embryo mass in the baseline case is $2\times 10^{-4}M_\oplus$, comparable to the mass of Ceres. This is also similar to the mass of the largest objects formed in recent simulations of planetesimal formation by the streaming instability \citep{schafer:2017}.

\begin{figure}
\plotone{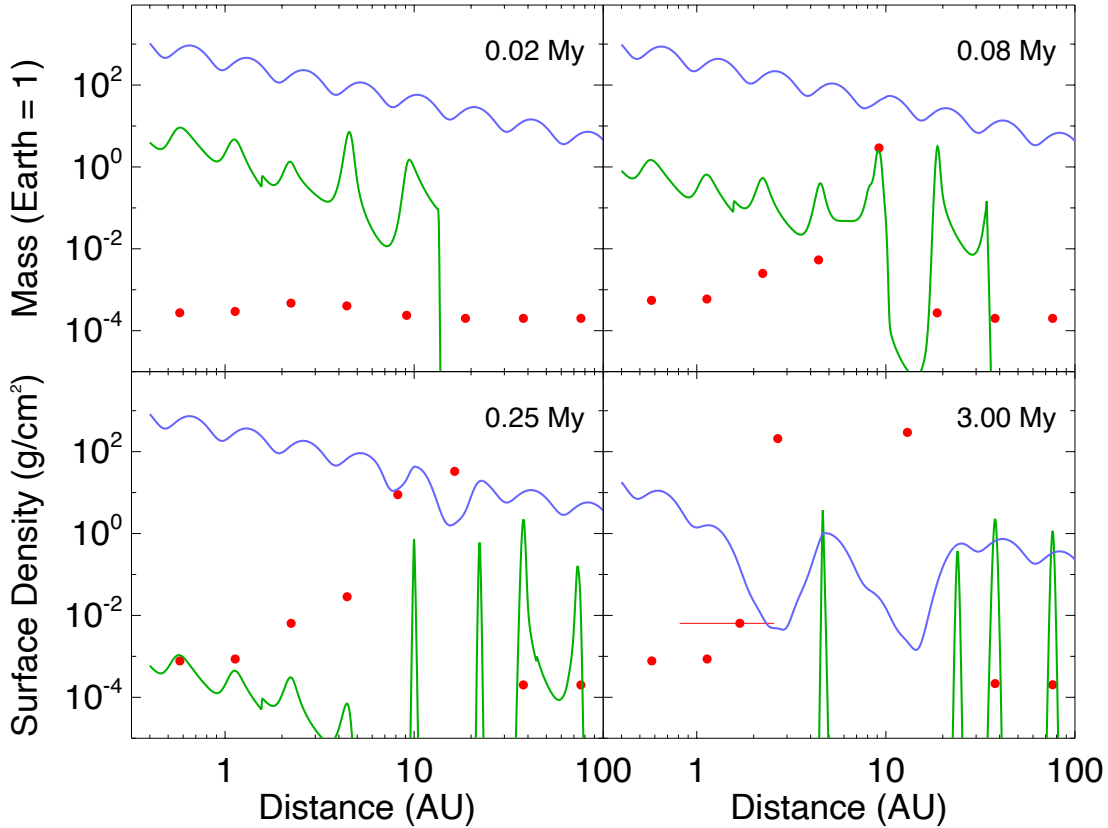}
\caption{Evolution in the baseline model at four times. The blue and green curves show the gas and pebble surface densities, respectively, as a function of distance. The red dots show embryo masses and locations. Red horizontal lines show the orbital extent for eccentric orbits.}
\end{figure}

%
%
\section{Results}
Figure 1 shows the evolution in a simulation using the baseline model. The model parameters in this case are given in Table~1. The gas disk has 8 surface density maxima, spaced logarithmically in distance with a semi-major axis ratio of 2 for neighboring bumps. We assume that a single planetary embryo has formed at each bump, located at the point where the pebble radial velocity is zero, if it exists, or at the point where the magnitude of the radial velocity is smallest. Pebbles are added at each radial zone in the disk after 400 orbital periods.

%
%
\subsection{Pebble Evolution}
The panels in the figure show the state of the system at 4 times. The blue curves show the gas surface density in g/cm$^2$. The series of sinusoidal bumps is clearly apparent in the first two panels of the figure. The gas surface density profiles in the last 2 panels are modified substantially as the 2 most massive embryos begin to open gaps in the disk.

Throughout the disk, pebbles tend to accumulate near the surface density maxima, forming bumps in the pebble surface density profile, which is shown in green. The pebble bumps are narrower than the gas bumps, but have a finite width set by a balance between radial drift and turbulence \citep{dullemond:2018}. Some transient features arise in the pebble surface density profile as the newly added pebbles adjust to the presence of the bumps in the gas disk. There is also an almost imperceptible jump in the surface density at the water ice line at 1.6 AU.

\begin{figure}
\plotone{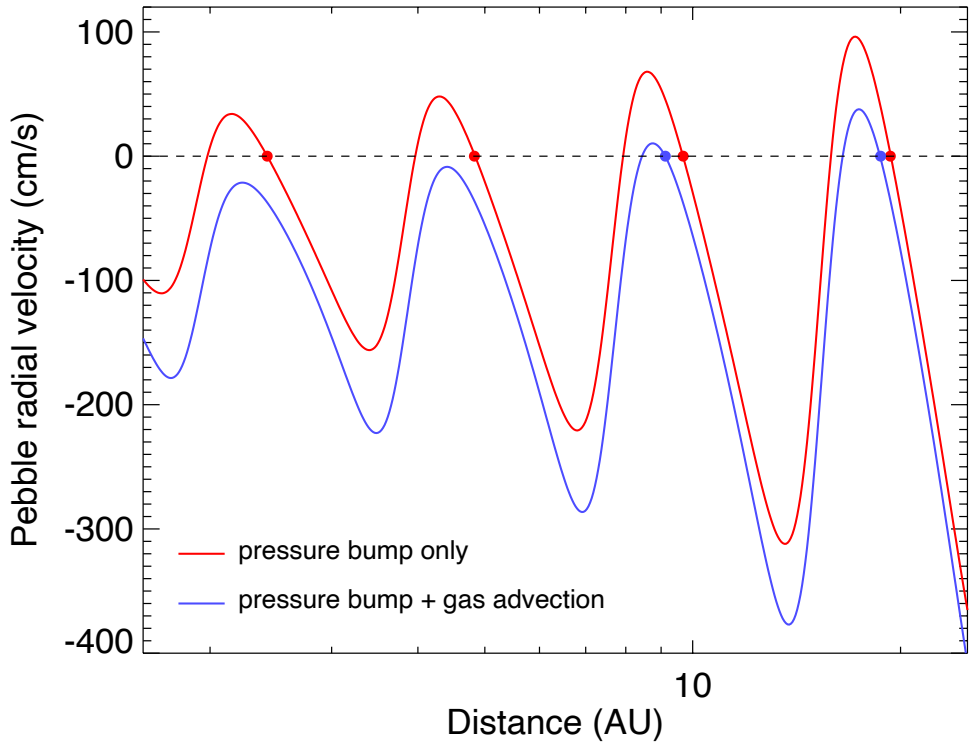}
\caption{Radial drift velocities of pebbles early in the simulation shown in Figure~1. The blue curve shows the actual radial velocities, with negative values indicating motion towards the star. The red curve shows the radial velocities that would exist if the inward advection of the gas were ignored. Stable equilibrium points (pebble traps) are indicated by circles. Note that traps only exist for some of the maxima in the blue curve.}
\end{figure}

Early in the simulation, the peaks in the pebble surface density profile in the outer disk typically have higher maxima than those in the inner disk. The reason for this difference can be understand by examining Figure~2, which shows the radial velocity of the pebbles early in the simulation before planets start to open gaps in the disk. The blue curve shows the actual radial velocities, with negative values indicating inward drift. Stable equilibrium points (``pebble traps'') are indicated by circles in the figure.

In the outer disk, each bump contains a single pebble trap where the radial velocity is zero. Pebbles accumulate in these locations as a result. This is not the case in the inner disk, where the peak values of the radial velocity are negative. Pebbles drift relatively slowly at these peaks, but they still drift. The peak surface densities are lower as a result. At later times, the inward flux of pebbles is greatly reduced, and the pebble bumps decay quite rapidly, as can be seen in the last 2 panels of Figure~1.

The absence of stable equilibrium points in the inner disk occurs because the radial velocity of the pebbles is set by a combination of the headwind (or tailwind) associated with the pressure gradient plus the inward advection velocity of the gas. In the inner disk, the  gas flows inwards too rapidly for pebble traps to exist. In the outer disk, radial velocities associated with the pressure gradient are larger (because the pebble Stokes number increases with distance), and pebble  traps do form, at slightly different locations than if gas advection were ignored. In the absence of gas advection, every bump would have a pebble trap, as shown by the red curve in Figure~2. 

\begin{figure}
\plotone{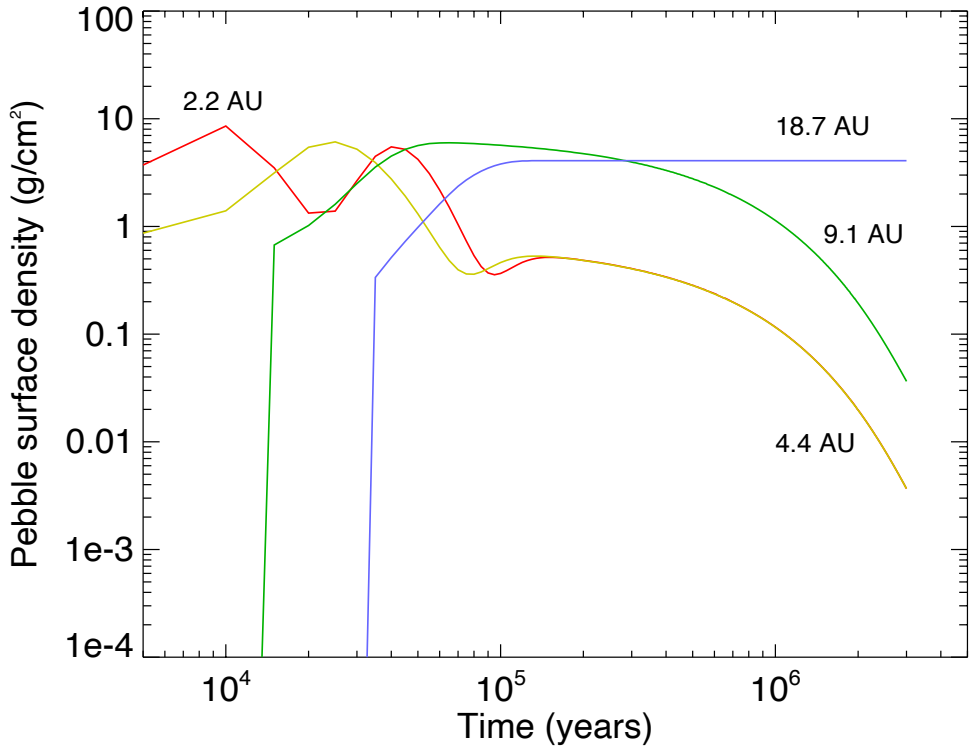}
\caption{Evolution of the pebble surface density at 4 bumps in a simulation with model parameters identical to Figures~1 and 2 except that planets are not included.}
\end{figure}

Figure 3 shows the evolution of the pebble surface density at the pressure bumps shown in Figure~2. Here, we consider the evolution of a system identical to Figure~1 except that protoplanets are absent. This allows us to gauge how pebble advection and diffusion affect the supply of pebbles available to the protoplanets in Figure~1.

In Figure 3, there are some initial transients as the pebbles adjust to the presence of the bumps in the gas disk. After about $10^5$ years, the surface density at each of the 4 bumps varies monotonically with time. The surface density at the outermost bump (blue curve) is essentially constant after the first $10^5$ years. Pebbles in this bump are trapped near a location where the radial velocity is zero. Pebbles at larger distances also become trapped at pressure bumps, so the source and sink terms for the bump at 19 AU both become negligible.

At the innermost two bumps (red and yellow curves), the pebble surface density decays monotonically after $10^5$ years due to the absence of a pebble trap.

The behavior at the third bump (green curve) in Figure~3 is somewhat similar. In this case a pebble trap does exist, which is why the surface density is an order of magnitude higher than the two bumps interior to it. However, the outward radial velocities in this bump are quite small (see the peak in the blue curve near 9 AU in Figure~2). As a result, some pebbles can diffuse across the bump. This generates a modest flux of pebbles that leaks into the region interior to the bump. Over time, the surface density at the bump near 9 AU declines as pebbles are lost and not resupplied from further out. The same flux supplies the bumps closer to the star, which is why the red and yellow curves track the green curve after the transients have died away.

%
%
\subsection{Embryo Evolution}
Returning to Figure~1, the masses and orbits of the embryos are shown by the red circles. The embryo growth histories are very different depending on their location in the disk. The embryos that  begin interior to 6 AU and exterior to 30 AU grow little, if at all, over the course of the simulation. In marked contrast, the embryos at 9 and 19 AU grow very fast, becoming massive enough to start accreting gas after only about 0.14 My. By 3 My, they have masses of 208 and 295 Earth masses, respectively, roughly comparable to Jupiter. 

The absence of pebble traps in the inner regions of the disk, and the resulting low pebble surface densities at the bumps in this region, helps explain why the embryos interior to 6 AU fail to form cores massive enough to accrete gaseous envelopes. In addition, the total mass of pebbles in the vicinity of each of these bumps is smaller than for bumps further from the star.

The growth of the two outermost embryos in Figure~1 also stalls, but for a different reason. Growth rates due to pebble accretion are sensitive to the pebble Stokes number $\st$, particularly close to the critical Stokes number $\stcrit$ as shown by Eqn.~\ref{eq-rcap}. The critical Stokes number itself depends sensitively on the relative velocity between pebbles and an embryo as shown by Eqn.~\ref{eq-stcrit}. 

\begin{figure}
\plotone{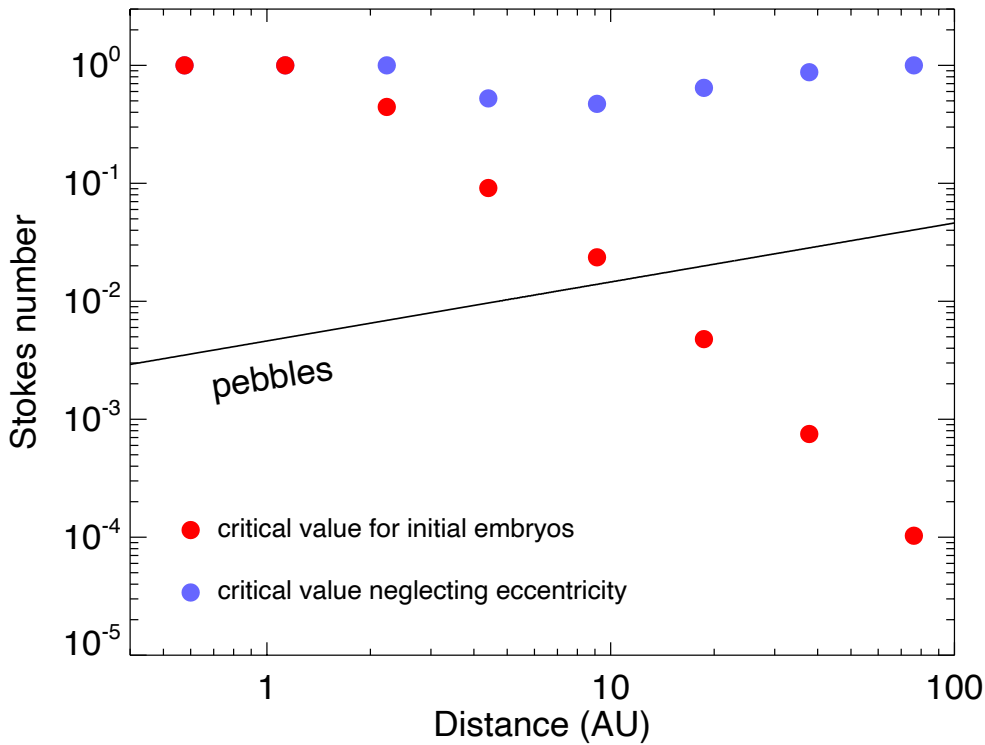}
\caption{Pebble Stokes number $\st$ (black curve) and critical Stokes number $\stcrit$ for pebble accretion (red circles) early in the simulation shown in Figure~1. Pebble accretion stalls when $\st\gg\stcrit$. The blue circles show the values of $\stcrit$ if the embryos were assumed to have circular orbits.}
\end{figure}

Figure 4 shows the critical Stokes number at the initial location of each of the embryos in Figure~1 (red, circular symbols), and the Stokes numbers of the pebbles as a function of distance (solid line). The pebble Stokes number increases slowly with distance while the critical value declines rapidly. 

The 6 innermost embryos have $\st$ smaller than or comparable to $\stcrit$. In these cases, pebbles are strongly affected by gas drag during an encounter with an embryo. Pebble accretion is efficient and embryos can grow rapidly as long as pebbles are abundant. Conversely, for the outer 2 embryos, $\st\gg\stcrit$. Pebbles encountering these embryos pass by too quickly for gas drag to significantly enhance the collision probability. Pebble accretion by gravitational focussing (see Eqn.~\ref{eq-rgrav}) is also ineffective in this region. As a result, these embryos grow extremely slowly.

The critical Stokes number decreases with distance because the embryo's eccentricities and relative velocities increase with distance. This occurs because excitation due to turbulent density fluctuations becomes increasingly important at larger distances (see Eqn.~\ref{eq-turb-excite}) compared to damping due to tidal torques. If the embryos were assumed to have circular orbits instead, the critical Stokes numbers  would be much larger (shown by the blue circles in Figure~4). The outer 2 embryos would grow much more rapidly as a result.

In Figure~1, the two embryos that grow into giant planets each undergo some inward migration. The amount of migration and the final masses of their gaseous envelopes are both limited by the fact that the embryos open deep gaps in the gas disk, which can be seen in the last panel of Figure~1.

The degree of inward migration is also moderated by the existence of a planet trap early in the planets' growth history. The combination of the Lindblad and unsaturated corotation torques mean that there is a particular location in each pressure bump where the radial migration rate is zero. This planet trap disappears when the planet grows more massive due to the weakening of the corotation torque (the exponential factor in Eqn.~\ref{eq-dadt}). 

\begin{figure}
\plotone{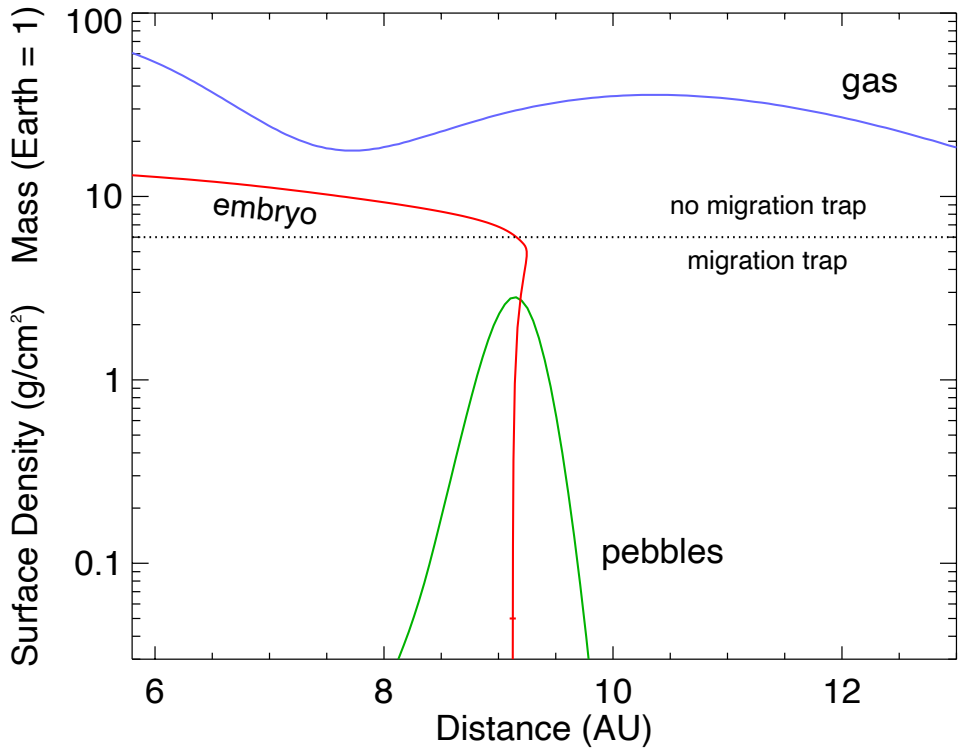}
\caption{The trajectory of one of the embryos in Figure~1 in mass-orbital distance space (red curve). Also shown are the gas (blue) and pebble (green) surface density profiles early in the simulation. A planet trap, where orbital migration ceases, exists slightly outside the embryo's initial location when its mass is less than 6 Earth masses.}
\end{figure}

This behavior can be seen more clearly in Figure~5, which shows the trajectory of the embryo located near 9 AU in mass-distance space (red curve). Also shown are the surface density profiles of the gas (blue) and pebbles (green) early in the simulation. A planet trap exists for embryo masses below about $6M_\oplus$ (see Appendix~C). 

The embryo begins somewhat interior to the planet trap because we have assumed it forms at the point where the pebble radial velocity is zero, and these two locations do not necessarily coincide \citep{morbidelli:2020}. As the embryo grows, it begins to migrate outward toward the planet trap. However, the growth rate is so fast that it never actually reaches the trap before the trap disappears. After that, the embryo migrates inward quite rapidly, reaching 2.7 AU at 3 My.

\begin{figure}
\begin{center}
\includegraphics[height=12cm]{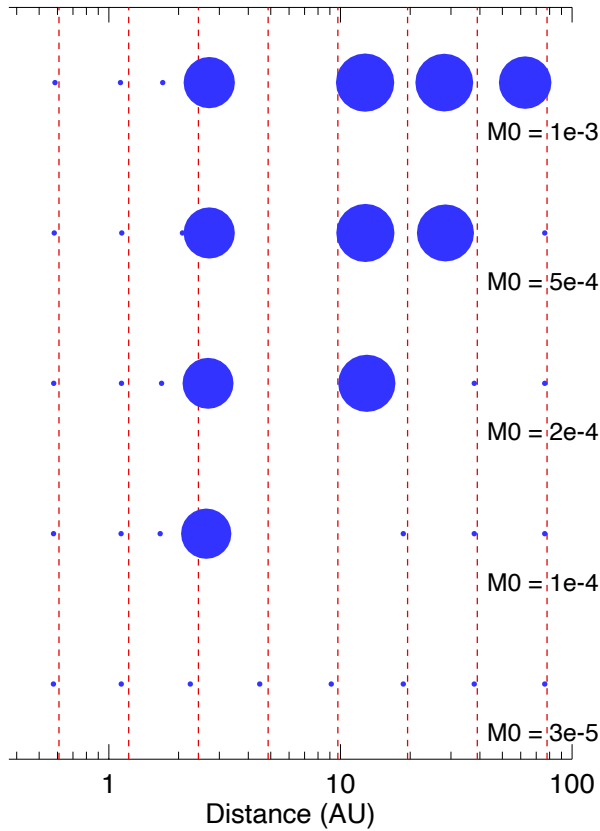}
\end{center}
\caption{The final planetary configurations for 5 simulations using different initial embryo masses $M_0$, given in Earth masses. Other parameters are identical to the case shown in Figure~1. Symbol radii are proportional to planetary radii assuming a fixed density. The red dashed lines show the locations of the pressure bumps.}
\end{figure}

%
%
\subsection{Dependence on Initial Embryo Mass}
Figure 6 shows the final outcome of 5 simulations each using a different initial embryo mass $M_0$. All other parameters are the same as the baseline case shown in Figure~1.

The outcome is clearly sensitive to the initial embryo mass. A factor of 30 difference in initial mass (3 in radius), is the difference between forming 4 gas giant planets or none at all. Increasing the initial embryo mass increases the number of giant planets that form. In most cases where an embryo fails to become a giant planet its mass remains close to the initial value, especially in the outer disk, leading to a pronounced dichotomy in outcomes.

The different outcomes are confined to the outer regions of the disk, in particular for the embryos that have initial orbits beyond 6 AU. The embryos closer to the star follow a similar evolution in each of the 5 simulations, except where they are strongly perturbed by more distant objects.

The different outcomes arise because of the sensitive dependence of pebble accretion growth rate on the value of the Stokes parameter compared to the critical value, seen in Eqn.~\ref{eq-stcrit}. The value of $\st$ at a particular distance is the same in each of the simulations for the model used here, but $\stcrit$ varies significantly from case to case. 

There are two reasons for this. Firstly, $\stcrit$ increases linearly with embryo mass. More importantly, $\stcrit$ depends on the inverse cube of the relative velocity between pebbles and an embryo. Less massive embryos experience weaker eccentricity and inclination damping due to tidal torques. The excitation of $e$ and $i$ by turbulent density fluctuations is independent of embryo mass. As a result, smaller embryos have larger velocities with respect to the pebbles. In some cases this can shut down pebble accretion almost entirely, leading to outcomes like the last planetary system shown in Figure~6.

\begin{figure}
\begin{center}
\includegraphics[height=12cm]{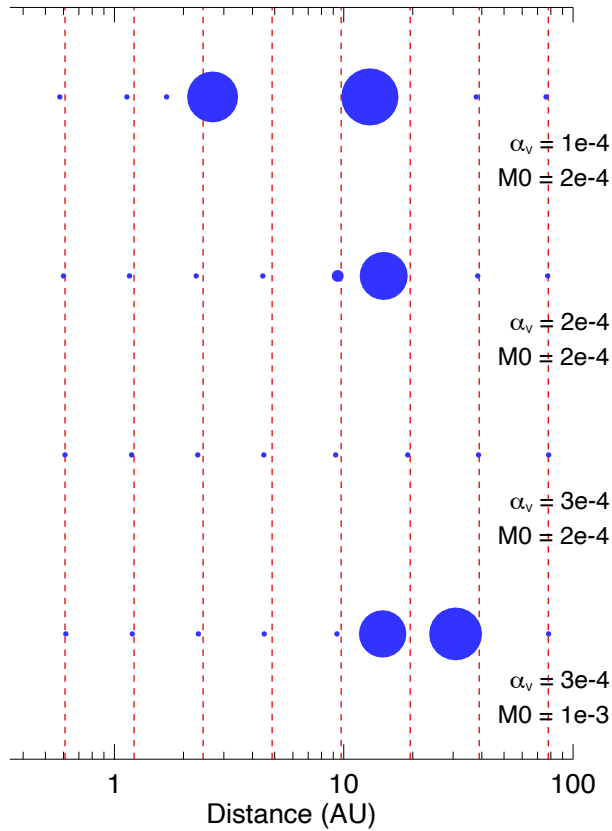}
\end{center}
\caption{The final planetary configurations for 4 simulations using different values of the turbulence parameter $\alpha_v$ and initial embryo mass $M_0$. Other parameters are identical to the case shown in Figure~1. Symbol radii are proportional to planetary radii assuming a fixed density. The red dashed lines show the locations of the pressure bumps.}
\end{figure}

%
%
\subsection{Dependence on Turbulence Strength}
The top three rows of Figure~7 show the effect of varying the strength of the turbulence in the disk gas, with other model parameters identical to the baseline case. Increasing the turbulence strength from $\alpha_v=10^{-4}$ to $3\times 10^{-4}$ reduces the number of gas giants that form from two to zero. 

Stronger turbulence adversely affects planetary growth for different reasons depending on the location in the disk. When the embryos are close to their initial mass, pebble accretion occurs in 3 dimensions, following Eqn.~\ref{eq-dmdt-peb}. In the inner region of the disk, the pebble Stokes number $\st$ is smaller than the critical value $\stcrit$ so that the capture radius $\rcap\simeq\rset$. Using the expressions for $\rset$ and $\vrel$ from Appendix A, and the relation between $\st$ and $\alpha_v$ given in Eqn.~\ref{eq-st}, we find that the growth rate in the inner disk follows
\begin{equation}
\frac{dM}{dt}\propto\frac{1}{\alpha_v^2}
\end{equation}

Thus, when $\alpha_v=3\times 10^{-4}$, the growth rate in the inner disk is an order of magnitude slower than the case with $\alpha_v=10^{-4}$. As a result, large embryos are unable to form in this region before the supply of pebbles decays away.

In the outer disk, pebbles remain available for much longer in the vicinity of the pebble traps. However, increasing the turbulence strength increases the relative velocity of pebbles with respect to the embryos because it makes the turbulent density fluctuations more effective at raising the embryo's eccentricity. Thus when $\alpha_v=3\times 10^{-4}$, we find that $\st\gg\stcrit$ in the outer disk.

Pebble accretion and planetary growth can still be effective in the outer disk if the initial embryos are more massive such that $\st$ is comparable to or smaller than $\stcrit$. The last row of symbols in Figure~7 shows an example. Here the turbulence parameter is $\alpha_v=3\times 10^{-4}$ as in the previous example, but the initial embryo mass has been increased by a factor of 5. In this case, two gas giants form in the outer disk, while growth in the inner disk is curtailed as before.

\begin{figure}
\begin{center}
\includegraphics[height=12cm]{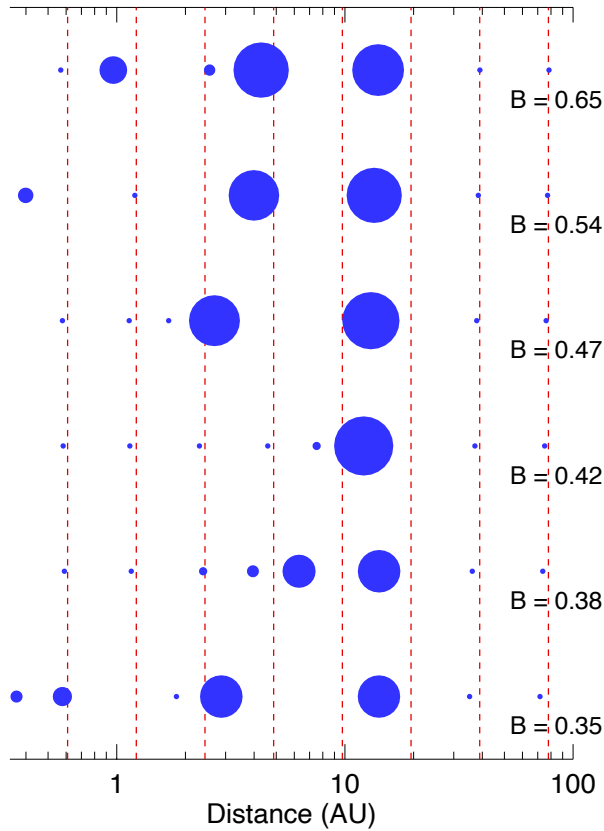}
\end{center}
\caption{The final planetary configurations for 6 simulations using different values for the pressure bump height $B$. Other parameters are identical to the case shown in Figure~1. Symbol radii are proportional to planetary radii assuming a fixed density. The red dashed lines show the locations of the pressure bumps.}
\end{figure}

%
%
\subsection{Dependence on Bump Height}
One aspect of the model that is especially uncertain is the magnitude of the bumps in real protoplanetary disks. Bumps may be more pronounced in some disks than others, and different bumps in the same disk could have different magnitudes. Here we briefly examine the importance of bump height assuming that all the bumps have the same relative height.

Increasing the bump height parameter, $B$, has two effects. Firstly it increases the height of positive bumps in the pebble radial velocity profile. Secondly, it can introduce additional positive bumps by raising maxima in the velocity profile above zero. For example, if $B$ is increased, some of the peaks in the blue curve in Figure~2 that lie below the dashed line can be raised above the line, creating traps where the radial velocity of pebbles is zero.

Figure~8 shows the outcome of 6 simulations with different values of the bump height, defined in Eqn.~\ref{eq-fbump}. The other model parameters are the same as the simulation shown in Figure~1. Unlike the initial embryo mass, the bump height mainly affects the growth of embryos in the inner portion of the disk. The growth of the outer two embryos stalls close to the initial mass in all of the simulations. 

In the baseline model, which has $B=0.47$, there are no pebble traps in the inner disk. This limits the supply of pebbles when pebble traps exist further from the star. As a result, the inner embryos remain small.

When the bumps in the gas disk have a larger amplitude, pebble traps exist can closer to the star than the baseline case. In the simulations with $B=0.54$ and $B=0.65$, the innermost pebble trap lies at 4.6 and 2.3 AU, respectively, compared to 9.1 AU in the baseline case. This increases the amount of pebbles available in the inner disk at early times. As a result, intermediate-mass planets are able to form (and migrate) in each case: a $33M_\oplus$ planet at 0.9 AU for the case with $B=0.65$, and a $6M_\oplus$ planet at 0.4 AU for the case with $B=0.54$.

However, an $11M_\oplus$ planet also forms in the inner disk in the simulation with the smallest value of $B$ shown in the last row of Figure~8. This shows that the effect of the bump height on the outcome is quite complex.

\begin{figure}
\plotone{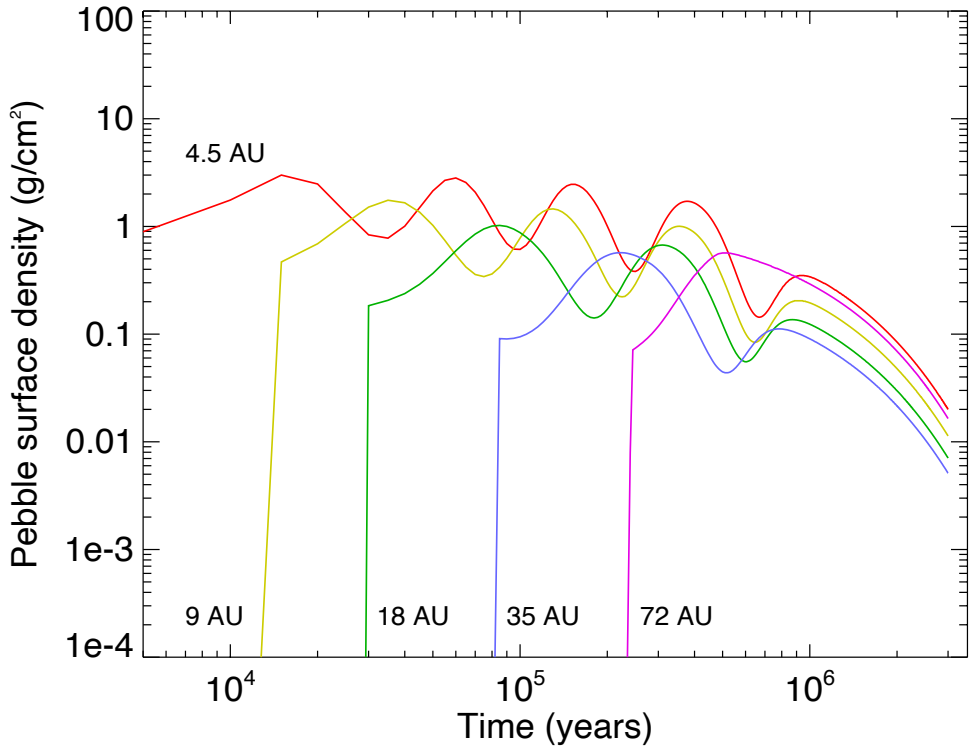}
\caption{Evolution of the pebble surface density at 5 bumps in a simulation with model parameters identical to the case with $B=0.35$ in Figure~8 except that planets are not included.}
\end{figure}

Much of this complexity arises because the pebbles take time to adjust to the presence of the gas bumps---typically 1--3$\times 10^5$ years. During this time interval, the pebble surface density profile undergoes substantial oscillations, as we saw in Figure~3. A favorably placed embryo can grow quite large in this timespan even if $B$ is too small to allow a pebble trap to exist at that location. 

The oscillations are especially long lived when $B$ is small because pebble traps only exist far from the star. Figure~9 shows an extreme example, corresponding to the case with $B=0.35$ in Figure~8, in which only a single pebble trap exists at 72 AU. In this example, the pebble surface density oscillates with time for roughly the first million years of the simulation.

A second factor, also apparent in Figures 3 and 9, is that some pebble traps are leaky. The outward radial velocities at the innermost pebble trap are typically quite small so that some pebbles can diffuse across the bump. This generates a flux of pebbles into the region interior to the innermost pebble trap, allowing embryos here to keep growing. Typically this flux is not sufficient to generate Jupiter-mass planets, but intermediate-mass planets can form in some cases.

Embryos can be much more effective at preventing the inward drift of pebbles than the intrinsic disk bumps in our model. Previous works have found that a large embryo can generate a pressure bump slightly outside its orbit and stop any pebbles flowing inwards \citep{lambrechts:2014a, bitsch:2018b, chambers:2018}. For example, Figure~10 shows how the presence of a $15M_\oplus$ planet located at 9 AU generates a much larger perturbation in the pebble drift velocity than the nearest intrinsic bumps. 

\begin{figure}
\begin{center}
\includegraphics[height=12cm]{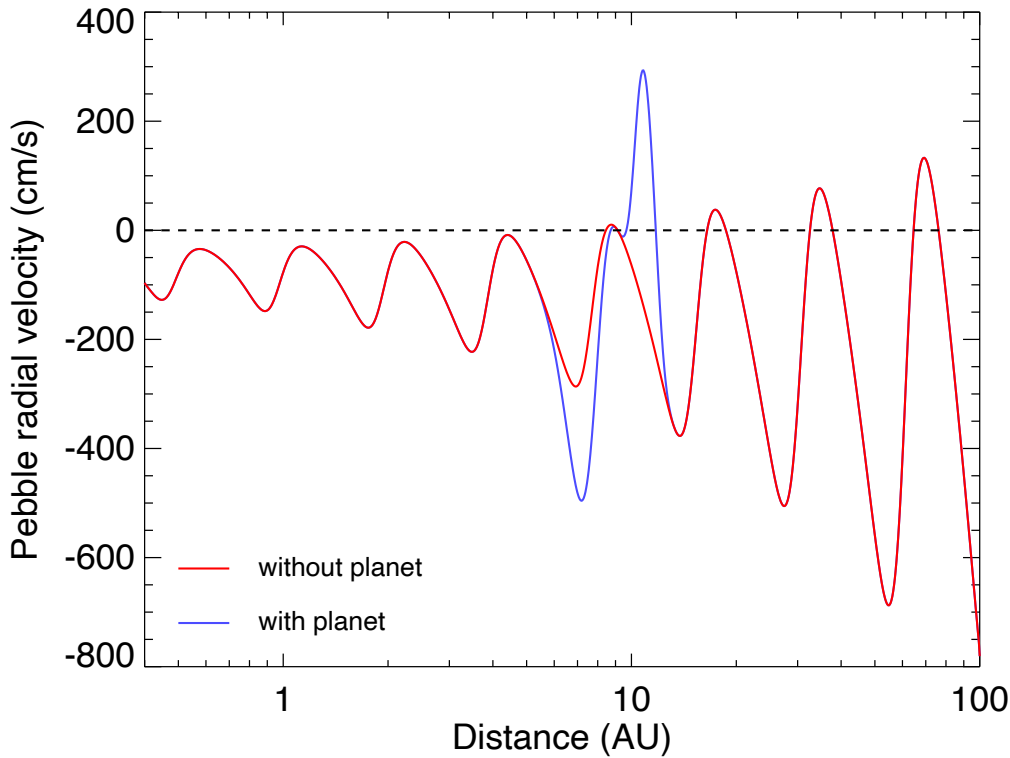}
\end{center}
\caption{Radial drift velocity of pebbles in a simulation without planets (red curve), and with a $15M_\oplus$ planet at 9 AU (blue curve). The planet generates a much larger positive deviation in the radial velocity than the intrinsic disk bumps, resulting in a more effective pebble trap.}
\end{figure}

The fate of planets in the inner disk can depend on the time at which a more distant planet becomes massive enough to shut off the supply of pebbles that would otherwise leak inwards. For example, in the baseline model shown in Figure 1, the formation of the 2 massive planets ultimately stops pebbles flowing inwards. However, the pebble traps caused by the intrinsic disk bumps also slow the growth of the inner planets at early times by reducing the amount of pebbles available. Thus, the final outcome depends on both these processes.

\begin{figure}
\begin{center}
\includegraphics[height=12cm]{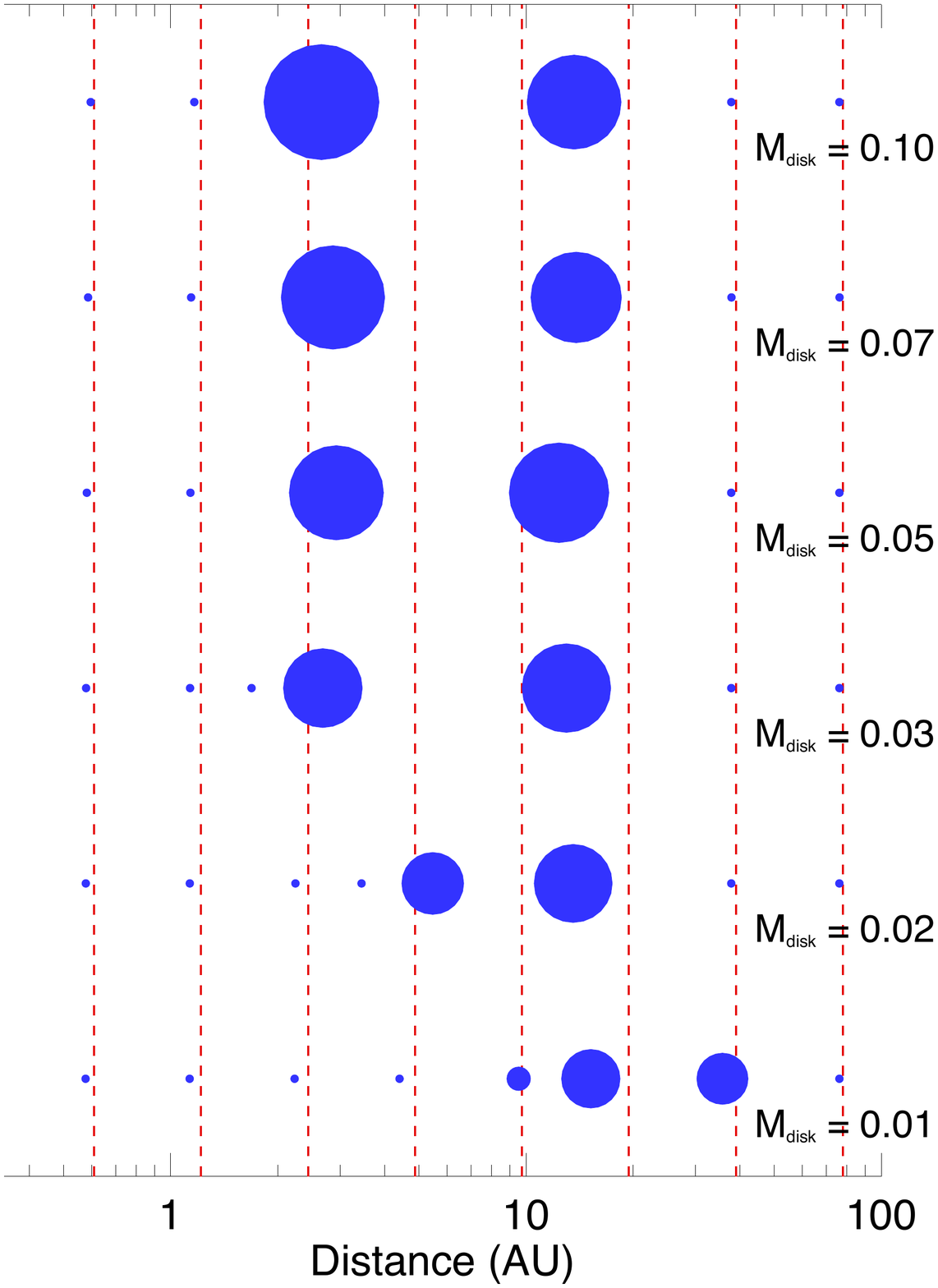}
\caption{The final configurations of 6 simulations using different initial disk masses $M_{\rm disk}$, given in solar masses. Other parameters are identical to the case shown in Figure~1. Symbol radii are proportional to planetary radii assuming a fixed density. The red dashed lines show the locations of the pressure bumps.}
\end{center}
\end{figure}

%
%
\subsection{Dependence on Disk Mass}
The baseline model considered here uses a relatively low disk mass of $0.03M_\odot$. Observed disks have a range of masses including ones that are more massive than the baseline case \citep{tychoniec:2018}. Here we explore the effect of disk mass on planet formation, focussing in particular on the formation of giant planets.

Figure~11 shows the final planetary systems formed in 6 simulations with different initial disk masses $M_{\rm disk,0}$=0.01--0.1 solar masses. The other model parameters are the same as the simulation shown in Figure~1.

For $M_{\rm disk,0}\ge0.03M_\odot$, changing the disk mass has a relatively modest effect on the outcome. These simulations all form 2 giant planets with roughly similar orbits in each case. The mass of the inner giant increases roughly linearly with the disk mass, while there is no clear trend for the mass of the outer giant.

The inner giant-planet core begins to form near the bump at 9 AU. It reaches the mass needed for rapid, hydrodynamic gas accretion at progressively earlier times with increasing disk mass. This is mostly because the core stops accreting pebbles at a higher mass so that the slow cooling phase of gas accretion is shorter. However, the tendency to open a deep gap in the gas disk prevents the earlier-forming cores from growing much more massive than those that form late. When viewed in mass-orbital distance space, the evolution follows a similar trajectory in each of these simulations.

The mass of the inner giant declines more rapidly with disk mass for $M_{\rm disk,0}\le0.03 M_\odot$. These planets experience less inward migration as a result. For the lowest disk mass considered here, the core at 9 AU remains too small to undergo rapid gas accretion before the simulation ends. In this case, the final planet mass is only about $5 M_\oplus$.

The outer gas giant in most of the simulations begins at the bump located at 19 AU. The planets in more massive disks actually grow more slowly than those in low-mass disks, at least initially. This is because pebble accretion at this location is prevented when the gas surface density $\sigmagas$ is too high due to excitation of the eccentricity, $e$, by turbulent density fluctuations. 

Eccentricity damping and excitation have different dependencies on $\sigmagas$ such that $e$ increases with initial disk mass. Efficient pebble accretion at 19 AU can only begin once the initial gas surface density has declined to a critical value, and this takes longer in high-mass disks.

As with the inner giant-planet core, the slow gas-cooling phase for the outer core is typically shorter in high-mass disks. This allows these planets to catch up with those formed in low-mass disks in some cases, leading to a complicated relation between disk mass and the final mass of the core that begins at 19 AU.

The dependence of $e$ on gas surface density, and thus $M_{\rm disk,0}$, leads to a qualitatively different outcome in the simulation with the lowest disk mass. (See the last row of symbols in Figure~11.) Here, a 58-Earth-mass planet forms at the bump near 38 AU whereas the corresponding embryos in the more massive disks undergo almost no growth. In the lowest-mass case, the gas surface density near 38 AU becomes low enough during the simulation for pebble accretion to begin. Thus, the low initial disk mass allows the embryo at this location to evolve into an intermediate-mass giant planet where no such object can form in more massive disks.

\begin{figure}
\plotone{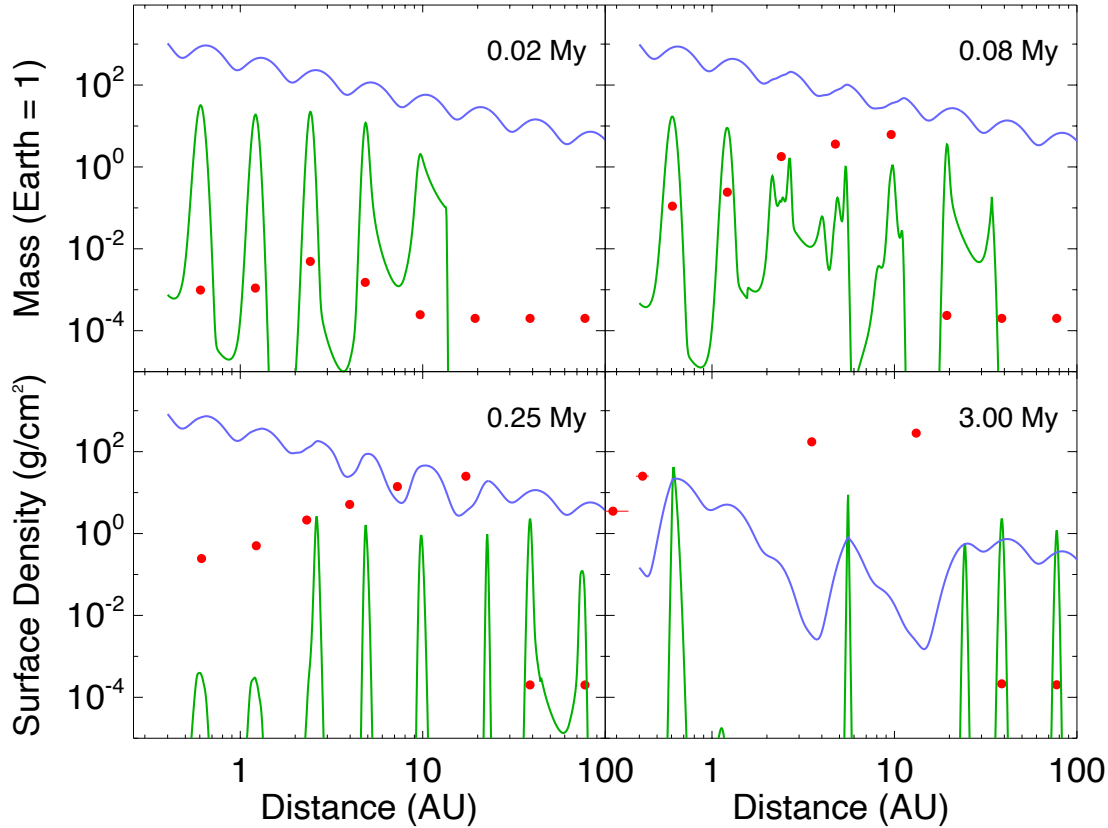}
\caption{Evolution when the radial drift of pebbles is assumed to be unaffected by the radial advection of the gas. Other model parameters are identical to the case shown in Figure~1. The blue and green curves show the gas and pebble surface densities, respectively, as a function of distance. The red dots show embryo masses and locations. Red horizontal lines show the orbital extent for eccentric orbits.}
\end{figure}

%
%
\subsection{Neglecting Pebble Drift due to Gas Advection}
Until this point, we have assumed that the radial drift of the pebbles is determined by both the local pressure gradient in the disk and the inward advection of the gas, following Eqn.~\ref{eq-vr}. It is possible, however, that the inward motion of the gas is largely confined to the surface layers of the disk. In contrast, the pebbles are concentrated closer to the disk midplane. This means that the effect of the gas advection on pebble drift may be less than we have assumed so far, and could even be negligible. Here, we examine this possibility.

Figure 12 shows a simulation with identical model parameters to Figure~1 except that the last term in Eqn.~\ref{eq-vr} is neglected. In this case, every pressure bump in the disk has a pebble trap where pebble advection ceases. The bumps are also higher than the baseline case shown in Figure~1, which makes the pebble traps more effective. Pebble diffusion across the pressure bumps is minimal as a result. Once the early transients have died away, the pebbles remain localized for the much of the simulation. The only exception occurs when pebble traps move as a result of planetary migration.

In the absence of migration, each planetary embryo can accrete mass only from the pressure bump where the embryo formed. The mass of pebbles available at each bump increases linearly with distance from the star, so more distant embryos grow larger. This can be seen in the second and third panels in Figure~12 before the onset of largescale migration. The condensation of water ice beyond 1.6 AU boosts the available solid mass by an additional factor of 2 for all but the two innermost bumps.

Three of the embryos (the ones beginning at roughly 5, 10 and 19 AU from the star) grow massive enough to accrete gas. The innermost of these grows much larger than the analogous object in the simulation shown in Figure~1, the result of the fact that pebbles in the inner disk remain close to the nearest pressure bump rather than drifting into the star. The three embryos closest to the star also grow larger than in the baseline case, although 2 are removed by subsequent events. 

Migration complicates the orderly growth trend seen early in the simulation. The innermost of the 3 large embryos migrates to the inner edge of the disk, sweeping up 2 of the smaller interior embryos in the process, a clear difference from Figure~1. However, the outer two large embryos finish with roughly similar masses and orbits to the case shown in Figure~1. This suggests that the effect of gas advection on the pebbles is more important in the inner disk than far from the star.

\begin{figure}
\plotone{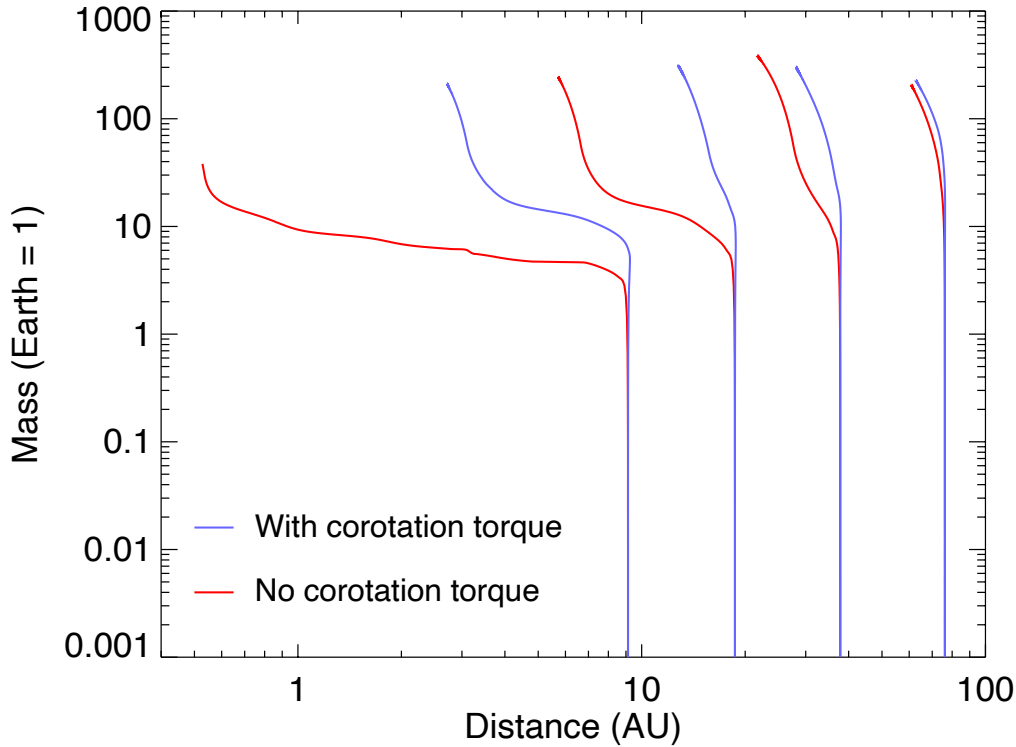}
\caption{Evolution of the four largest planets in mass-orbital distance space in a simulation with fully unsaturated corotation migration torques (blue curves), and a case that neglects corotation torques (red curves). The initial embryo mass is $10^{-3}M_\oplus$. Other model parameters are identical to the case shown in Figure 1.}
\end{figure}

%
%
\subsection{Neglecting Corotation Migration Torques}
In the simulations above, we have assumed that interactions between a planet and the gas disk include a corotation torque with the maximum strength described by \citet{paardekooper:2010}. The migration rate in this case is given by Eqn.~\ref{eq-dadt}. However, it is believed that the corotation torque can saturate in some circumstances so that the torques cease after a certain time \citep{paardekooper:2011}. In this section, we examine the effect of turning off the corotation torque on a planet leaving only the torques associated with Lindblad resonances.

Figure 13 compares the evolution in two simulations, one that includes corotation torques (blue curves) and one that neglects them entirely (red curves). The migration behavior of real planets is likely to lie somewhere between these extremes. The curves show the evolution in mass-orbital distance space for the 4 largest planets in a system with model parameters identical to Figure~1 except that the initial embryo mass is $10^{-3}M_\oplus$. (The blue curves correspond to the planetary system shown at the top of Figure~6.)

In both simulations, planets migrate inwards. However, the extent of migration is greater when corotation torques are neglected. When corotation torques are included, migration can slow or cease for low-mass planets as we saw in Figure~5. These planet traps disappear when the planet reaches a critical mass. However the existence of a trap early in a planet's growth typically reduces the total distance it migrates. 

Planet traps have a marked effect in the model used here because massive planets readily open deep gaps in the disk so that they also migrate slowly. Thus, there is a relatively narrow range of masses during which a planet can undergo rapid inward migration (the nearly horizontal sections of some of the curves in Figure~13).

With or without planet traps, migration in Figure~13 appears to be more important for planets closer to the star. Part of this apparent trend is the result of the logarithmic distance scale used in the figure---the migration distances measured in AU differ by only a bit more than a factor of 2 for each simulation. This is not too surprising since the scaling factor for the migration rate, given by Eqn.~\ref{eq-dadt0}, is independent of distance in the model used here. Nevertheless there is a weak trend towards greater migration at smaller distances. 

Finally, we note that in either of the two migration models, a gas giant planet can form at a distance of 60 AU, within 3 My, thanks to the concentration of pebbles at a pressure bump. 

%
%
\section{Discussion}
%
%
\subsection{The Difficulty of Forming Giant Planets}
Previous studies have found that it is hard to form systems of giant planets beyond 5--10 AU from a star by either planetesimal accretion or gravitational scattering between planets. If planetesimals are large (10--100 km), growth rates are too slow to form giant-planet cores before the star's protoplanetary disk dissipates \citep{johansen:2019}. If the planetesimals are smaller, they tend to get isolated in regions where they cannot be accreted due to a combination of gravitational scattering and gas drag \citep{levison:2010}. Planet-planet scattering seems to have a very low probability ($\sim 0.1$\%) of forming stable, multi-planet systems with wide orbits like HR 8799 \citep{dodson-robinson:2009}.

Alternatively, giant planets on wide orbits could form by gravitational instabilities in the disk. Numerical simulations by \citet{boss:2006} found that forming giants at 100--200 AU is unlikely, but simulations at 30--70 AU succeeded in forming gravitationally bound clumps that could be giant-planet precursors \citep{boss:2011}. The results of such simulations appear to depend sensitively on the algorithm used \citep{meru:2011, deng:2017}, which leaves open the question of whether some giant planets form by disk instability. However, an analysis of the observed size distribution of wide-orbit giants suggests that disk instability is probably not the dominant formation mechanism \citep{wagner:2019}.

Given these issues, much recent work on giant planet formation has focussed on pebble accretion \citep{ormel:2010, lambrechts:2012, bitsch:2015, matsumura:2017}. However, pebble accretion is not without problems. In disks with monotonic pressure gradients, pebbles can drift inwards rapidly \citep{weidenschilling:1977a}. Typically only a small fraction are captured by a planetary embryo before they have drifted past its orbit \citep{guillot:2014, morbidelli:2015}. This implies that a massive disk is required to supply enough pebbles to form a giant planet. A second issue is that disks with modest radii can lose all their pebbles in much less than the lifetime of the gas disk, leaving relatively little time to form giant planets \citep{bitsch:2018a}.

%
%
\subsection{Disks with Pressure Bumps}
As noted by \citet{morbidelli:2020}, both these problems with pebble accretion can be avoided if pebbles become trapped at long-lived pressure bumps in a protoplanetary disk, provided that enough pebbles are swept up into the trap to build a giant-planet core.

Using an analytic model, \citet{morbidelli:2020} found that a 10-Earth-mass core can form at a pressure bump 5 AU from a solar-mass star in about 0.3 My. \citet{guilera:2020} found even faster growth, with a $13 M_\oplus$ core forming in only 0.02 My at a bump located 3 AU from its star. This core subsequently accreted gas, reaching Saturn mass in 0.3 My.

However, \citet{morbidelli:2020} calculated much slower growth rates at distances comparable to some directly-imaged planets. For example, under the most optimistic circumstances he considered, a $0.1 M_\oplus$ core located at 75 AU grew to only about an Earth mass in 3 million years.

In contrast, we find much faster growth rates at large distances, provided that an embryo exceeds a critical mass initially. For the case shown in Figure~13, an embryo at 75 AU grows from $10^{-3}M_\oplus$ to $10M_\oplus$ in only a million years. The onset of gas accretion allows the embryo to reach $200M_\oplus$ by 2 My.

The growth rates calculated by \citet{morbidelli:2020} are limited mainly by the rate at which pebbles can move radially within a bump  to the location of the planetary embryo. Regardless of how efficiently an embryo accretes pebbles, its growth rate is throttled by the flow of pebbles towards it. The growth rate is also limited in some cases because the embryo was initially placed at the point where the migration rate was zero, some distance away from the peak of the pebble surface density.

There are several reasons why the supply of pebbles may not limit growth rates in our simulations. Firstly, the embryos in our model have eccentric orbits. This, coupled with the fact that an embryo accretes pebbles from a substantial fraction of its Hill radius, means that the embryo's feeding zone occupies a significant fraction of the width of the pressure bump.

Secondly, embryos in our model begin at the point where the pebble radial velocity is zero (the pebble trap) rather than the location where the embryo's migration rate is zero (the planet trap). The seed embryos considered in Section~3 have masses that are too low to be significantly affected by migration so there is no reason they should form at the planet trap. Instead, it makes more sense that an embryo would tend to form where the surface density of pebbles is greatest.

Finally, as an embryo grows it migrates, typically outwards at first in our model, towards the planet trap, and then inward at later times. This gives a growing embryo access to a larger portion of the pressure bump than if there were no migration. Migration enhances the effective feeding zone of the embryo. For all these reasons, we suggest that planetary growth rates may be larger than those calculated by \citet{morbidelli:2020} at large distances from the star.

%
%
\subsection{The Importance of Turbulent Density Fluctuations}
The effectiveness of pebble accretion in the outer disk is strongly influenced by turbulent density fluctuations. Turbulence creates over-dense regions in the gas that gravitationally scatter embryos, increasing their orbital eccentricities as a result \citep{kobayashi:2018}. This raises the encounter velocity between an embryo and the pebbles, reducing the chance that gas drag can slow pebbles enough to allow capture. This process is particularly important for small embryos in the outer disk because turbulent excitation becomes increasingly effective with distance compared to eccentricity damping by tidal torques. 

It has been known for some time that turbulent density fluctuations can limit runaway growth of planetesimals by increasing their velocity dispersion \citep{ida:2008, ormel:2013, kobayashi:2016}. In addition, \citet{rosenthal:2018} have noted that turbulence can increase the velocity dispersion of pebbles making them harder to capture. Here we note a new effect: turbulence increases the random velocities of the embryos too. This slows pebble accretion, sometimes dramatically (as seen in Figure~6, for example). 

It is worth pointing out that studies of pebble accretion often assume the embryos have circular orbits. If excitation by turbulent density fluctuations is important in real disks this simplification may lead to incorrect predictions, especially if the initial planetary embryos have radii $<1000$ km.

%
%
\subsection{Formation of Solar System Analogs}
Gas giant planets form very rapidly in the simulations presented here. Analogs of Jupiter appear within a million years in many cases. This implies Jupiter could have been present from an early stage during the formation of the Solar System. It is plausible that Jupiter significantly affected the dynamical and chemical evolution across much of the planetary system as a result.

In the baseline model, giant planets tend to form at intermediate distances, while objects close to the star and at large distances tend to remain small. We saw above how the growth of embryos in the outer disk can stall due to the effects of turbulent density fluctuations if the initial embryos are too small. 

Growth in the inner disk is limited for different reasons. Firstly, the total mass of pebbles in the vicinity of a pressure bump increases with distance from the star, so the innermost bumps do not contain enough pebbles to form a giant-planet core. Secondly, pebbles are readily lost from the inner bumps. In many cases no pebble trap exists in these bumps because the inward advection velocity of the gas is too high. Even when a trap does exist it can be ``leaky'' because pebbles can cross the bump via turbulent diffusion.

The tendency of giant planets to form at intermediate distances agrees with some of the observed features of the Solar System, in particular the absence of giants interior to 5 AU, and the marked contrast between the masses of the giant planets and the members of the Kuiper belt.

However, there are some important differences. The final masses of objects in the inner disk are typically lower than those of the terrestrial planets. We suggest this is because planetesimal accretion was an important factor in the growth of the inner planets, and this was neglected here.

We also note that analogs of Uranus and Neptune do not form in our simulations. Forming embryos with masses comparable to the cores of these planets at 15--30 AU is easily achieved in many of the simulations. However, these objects typically go on to accrete massive gaseous envelopes. This implies that some additional physics limiting the growth of these planets is missing in the model.

%
%
\subsection{Formation of Wide-Orbit Planets}
In Section~3 of this paper, we showed that gas giant planets can form rapidly by pebble accretion in a disk that has pressure bumps. Depending on the model parameters, giant-planet formation can be restricted to intermediate distances from the star, or giants can form in most regions of the disk, out to at least 60 AU. This is comparable to the orbital distances of some directly-imaged planets \citep{marois:2008, haffert:2019}.

The outcome of planet formation in the outer disk is sensitive to the initial planetary embryo mass, with a strong dichotomy between giant planet formation and essentially no growth. Models for planetesimal formation by the streaming instability and turbulent concentration both predict a spectrum of initial embryo masses, with the probability density declining at large masses \citep{cuzzi:2010, simon:2017}. This implies that chance differences in the initial embryos will lead to very different outcomes in different systems. This may explain why giant planets with wide orbits can exist but are relatively rare \citep{nielsen:2019}.

One obvious difference between the observed wide-orbit planets and those produced in the simulations is that the former are typically more massive by up to an order of magnitude \citep{nielsen:2019}. The model used here for gas accretion onto embryos is rather simplistic, especially for the regime in which the gas accretion rate is limited hydrodynamically. Recent work suggests that reality is a good deal more complex than we consider here, with three dimensional flows and circumplanetary disks strongly influencing the gas accretion rate \citep{szulagyi:2014}. As a result, there is much uncertainty regarding the asymptotic mass of gas-giant planets \citep{ginzburg:2019}. The final planetary masses found here should be treated with caution for this reason.

We note that simulations of giant planet formation often appeal to very massive disks in order to overcome inefficient growth and long growth timescales. In contrast, the initial disk mass used in the simulations in this paper is quite modest---0.03 solar masses---within the range of estimates for the minimum mass solar nebula \citep{weidenschilling:1977b, kuchner:2004, desch:2007}.  In fact, we find that planets exceeding $100M_\oplus$ can form within 2 My in the outer regions of a disk with a mass of only $0.01M_\odot$. This is a testimony to the efficiency of planet formation when pebble traps exist. The existence of such traps may allow giant planet formation in a wider range of protoplanetary disks than considered previously.

%
%
\subsection{Model Limitations}
The simulations presented in this paper employed a number of approximations. Here we discuss the likely impact of some of these simplifications.

The simulations in Section 3 considered growth due to pebble accretion and gas accretion. However, planetesimal accretion was neglected. The main emphasis of this work is to understand the origin of giant planets. As discussed above, planetesimal accretion is unlikely to be a major pathway to giant-planet formation, so neglecting planetesimals is probably a reasonable first approximation. However, planetesimal accretion is probably important for the formation of terrestrial planets and super Earths. This may explain why such planets are rarely produced in the simulations described here.

Ongoing planetesimal accretion may also delay gas accretion onto a planetary embryo even when the accretion of pebbles has stopped. Heat released by infalling planetesimals can prevent a giant-planet core from cooling and accreting gas in the same way that infalling pebbles do. For example, in the simulations by \citet{guilera:2020} described above, adding planetesimal accretion increased the time required to form a Saturn-mass planet at 3 AU from 0.3 My to roughly 1.5 My. However, that the importance of planetesimal accretion declines rapidly with distance from the star \citep{morbidelli:2015}. As a result, the formation of gas giants in the outer disk is unlikely to be altered substantially by planetesimal accretion.

A potentially more serious shortcoming is that we considered only a single planetary embryo at each pressure bump in the disk. Simulations by \citet{kretke:2014} find that when many embryos are present in a region, mutual perturbations and competition to accrete pebbles can lead to a situation similar to oligarchic growth of planetesimals. This can limit growth rates and lead to the formation of more giant-planet cores than we see in the Solar System for example.

Examining the importance of multiple embryos per pressure bump is beyond the scope of the present work. However, we note that perturbations due to  turbulent density fluctuations (not included in \citet{kretke:2014}) make the growth rate extremely sensitive to embryo mass in our model. Given that embryos are likely to form with a spectrum of initial masses, it is conceivable that one embryo will quickly grow much larger than its rivals in the same pressure bump as a result.

The model used here neglects the effect of partial gaps caused by planets on the radial velocity of the gas. It is plausible that the inward velocity of the gas increases in the vicinity of a gap in order to maintain the same inward mass flux. Although this effect is ignored, we note that gap opening tends to be important after most pebble accretion has already occurred, so this is unlikely to change the outcome greatly.

Finally, the simulations neglect the back reaction of pebbles on the gas due to drag forces. This effect is likely to become important when the pebble surface becomes comparable to the gas surface density. This occurs late in some of the simulations presented here, for example in the final panels of Figures~1 and 11. The distribution of pebbles is probably unrealistic in these cases. However, we note that most of the growth by pebble accretion occurs at a much earlier stage in the simulations, so this approximation may not be important in practice.

%
%
\section{Summary}
In this paper we present a model  for planet formation in a protoplanetary disk that features a series of pressure bumps. The bumps are assumed to be persistent, stationary, intrinsic features of the disk, logarithmically spaced in distance from the star. We consider the growth and orbital evolution of planetary embryos at each bump, subject to pebble accretion, gas accretion, damping and orbital migration due to tidal torques, and excitation due to turbulent density fluctuations. Planetesimal accretion is neglected.

The main results of this study are
\begin{enumerate}
\item Under suitable circumstances, Jupiter-mass planets can form out to at least 60~AU in less than 3~My in a 0.03 solar mass disk orbiting a solar mass star.

\item In the outer disk, pebbles drift rapidly towards the nearest pressure bump. Each bump has a pebble trap where the radial drift velocity is zero. Planets can grow rapidly at these locations.

\item In the inner disk, pressure bumps typically do not have a pebble trap because the inward advection velocity of the gas is too high. Pebbles are lost from the inner disk as a result.

\item Planetary growth rates are very sensitive to the initial embryo mass and the strength of turbulence in the disk. The outcome also depends on the relative height of the pressure bumps. Modest differences in these quantities in different protoplanetary disks can lead to very different outcomes.

\item In the outer disk, planetary embryos with initial masses below a (distance-dependent) critical value grow extremely slowly. This is because pebble accretion is frustrated by the fact that embryo orbits are eccentric due to turbulent density fluctuations in the gas.

\item Embryos above the critical mass rapidly become large enough to accrete gas, forming gas giant planets in as little as 0.5~My. This produces a marked dichotomy: most embryos either grow into gas giants or remain close to their initial mass.

\item Massive planets open deep gaps in the gas disk. This prevents the formation of planets significantly more massive than Jupiter, and it also moderates the extent to which the planets migrate.

\item Neglecting corotation migration torques or the effect of gas advection on pebble drift increases the formation rate and migration of super-Earths in the inner disk. These two processes have a minor effect on the outcome in the outer disk.
\end{enumerate}

\acknowledgments
I thank an anonymous reviewer for his or her comments that helped improve this paper.

%
%
\appendix

\section{Pebble Capture Radius and Relative Velocity}
Here we give expressions for the pebble capture radius and the embryo-pebble relative velocity used to calculate the pebble accretion rate. These expressions are based on \citet{ormel:2010}.

In the regime where pebbles are strongly affected by gas drag, and settle towards an embryo during an encounter, the capture radius is
\begin{equation}
\rset=r_H\times\min\left[
\left(\frac{12r_H\st}{a|\eta|}\right)^{1/2},
\left(12\st\right)^{1/3}\right]
\end{equation}
where $\st$ is the pebble Stokes number, $r_H$ is the embryo's Hill radius, and $\eta$ is the fractional amount by which the local orbital velocity of the gas $\vgas$ differs from the Keplerian velocity $\vkep$, where
\begin{equation}
\eta=\frac{\vgas-\vkep}{\vkep}
\end{equation}

The two terms in square brackets in the equation for $\rset$ refer to regimes in which the embryo affects the motion of pebbles in part or all of its Hill sphere respectively. 

The relative velocity between the embryo and an incoming pebble is given by
\begin{equation}
\vrel=\max\left[|\eta|\vkep,\frac{\rset\vkep}{a}
\right]
\end{equation}
where the two terms in square brackets refer to the same two regimes.

Note that this equation for $\vrel$ assumes the embryo moves on a circular orbit in the disk midplane. In our model, we use a more general expression for the relative velocity, given by Eqn.~\ref{eq-vrel}.

%
%
\section{Gas Accretion with Ongoing Pebble Accretion}
In this appendix, we describe how the gas accretion rate onto an embryo is modified when the embryo is also accreting pebbles. 

Gas and pebble accretion both release gravitational energy into an embryo's atmosphere counterbalancing the energy radiated away by the atmosphere \citep{lee:2015}. The atmosphere's internal energy also increases as its temperature rises. Here, we assume these processes balance. Thus, pebble accretion reduces the rate at which an embryo can accrete gas. If the pebble accretion rate is high enough, there will be no gas accretion.

At equilibrium, the energy of the atmosphere $E$ changes at a rate given by
\begin{equation}
\frac{dE}{dt}=-L+\lpeb
\end{equation}
where $L$ is the luminosity of the atmosphere, and $\lpeb$ is the energy released by infalling pebbles. We assume that the pebbles sediment to the core so that $\lpeb$ is given by
\begin{equation}
\lpeb=\frac{GM_c}{r_c}\left(\frac{dM_c}{dt}\right)_{\rm peb}
\label{eq-lpeb}
\end{equation}
where the subscript $c$ refers to the core of the embryo.

In most cases, the atmosphere of an embryo will be centrally concentrated, with most of the mass contained in an inner convective region surrounded by a low-density, outer radiative region \citep{lee:2015}. We assume that the density $\rho$ in the convective region can be approximated by a power law:
\begin{equation}
\rho=\rho_b\left(\frac{r}{r_b}\right)^{-n}
\end{equation}
where $n>3$ and the subscript $b$ refers to the radiative-convective boundary.

The mass and gravitational potential energy of the atmosphere are approximately given by
\begin{eqnarray}
\matmos&\simeq&4\pi\int_{r_c}^{r_b}\rho r^2\, dr \\
PE&\simeq&-4\pi GM_c\int_{r_c}^{r_b}\rho r\, dr 
\end{eqnarray}
The second expression assumes that the mass of the atmosphere is comparable to or smaller than the core mass, which is likely to be the case when pebble accretion is significant.

Inserting the equation for density, we get
\begin{eqnarray}
\matmos&\simeq&\frac{4\pi\rho_br_b^nr_c^{3-n}}{(n-3)}
\\
PE&\simeq&-\frac{4\pi GM_c\rho_br_b^nr_c^{2-n}}{(n-2)}
=-\left(\frac{n-3}{n-2}\right)
\frac{GM_c\matmos}{r_c}
\end{eqnarray}

The total energy of the atmosphere (gravitational plus internal) is
\begin{equation}
E=f\times PE=-f\times\left(\frac{n-3}{n-2}\right)
\frac{GM_c\matmos}{r_c}
\end{equation}
where we assume that $f$ is fixed, which is true for an ideal gas \citep{piso:2014}.

Thus
\begin{equation}
\frac{dE}{dt}\simeq
-f\times\left(\frac{n-3}{n-2}\right)
\frac{GM_c}{r_c}\frac{d\matmos}{dt}
\end{equation}
where we have neglected the energy change of the atmosphere due to changes in the core mass and radius.

Returning to the energy balance equation, we get
\begin{equation}
\frac{d\matmos}{dt}\simeq
\left(\frac{n-2}{n-3}\right)\frac{r_c}{GM_cf}[L-\lpeb]
\end{equation}

In the absence of pebble accretion, we use the gas accretion rate calculated by \citet{bitsch:2015}, given by Eqn.~\ref{eq-dmdtcool}, so that:
\begin{equation}
\frac{d\matmos}{dt}=
\left(\frac{n-2}{n-3}\right)\frac{r_c}{GM_cf}L
=\left(\frac{d\matmos}{dt}\right)_{\rm cool}
\end{equation}

We assume that the atmosphere structure satisfies $n=3.5$ and $f=0.2$. Using these values and Eqn.~\ref{eq-lpeb}, the general expression for the gas accretion rate in the presence of pebble accretion rate is
\begin{equation}
\frac{d\matmos}{dt}=\max\left[0,
\left(\frac{d\matmos}{dt}\right)_{\rm cool}
-15\left(\frac{dM_c}{dt}\right)_{\rm peb}\right]
\end{equation}
where $(dM/dt)_{\rm peb}$ is the pebble accretion rate.

Note that there is no gas accretion unless the pebble accretion rate is much smaller than the maximum possible gas accretion rate $(d\matmos/dt)_{\rm cool}$.

%
%
\section{Location of Migration Traps}
In this appendix, we derive the necessary condition for a planet migration trap to exist at a pressure bump.

The semi-major axis of a planet migrates at a rate given by Eqns.~\ref{eq-dadt}--\ref{eq-phi-beta}, which we combine into a single expression here
\begin{equation}
\frac{da}{dt}=
\left(\frac{da}{dt}\right)_0\left[
(2.05e^{-K/20}-3.35)
+\phi(0.1-1.1e^{-K/20})
\right]
\end{equation}
where
\begin{eqnarray}
\left(\frac{da}{dt}\right)_0&=&2\left(\frac{M}{\mstar}\right)
\left(\frac{\sigmagas a^2}{\mstar}\right)
\left(\frac{a}{\hgas}\right)^2\vkep
\nonumber \\
\phi&=&-\frac{d\ln\sigmanogap}{d\ln a}
\nonumber \\
K&=&\left(\frac{M}{\mstar}\right)^2
\left(\frac{a_p}{\hgas}\right)^5
\frac{1}{\alpha_v}
\end{eqnarray}
and we have incorporated the radial temperature profile used in our model.

For zero migration, $da/dt=0$, which requires that
\begin{equation}
K=20\ln\left(\frac{2.05-1.1\phi}{3.35-0.1\phi}\right)
\label{eq-ktrap}
\end{equation}
Note that the largest values of $K$ occur when $\phi$ is negative.

In our model for bumps, the gas surface density is given by
\begin{equation}
\sigmanogap=\Sigma_0\left(\frac{a}{a_0}\right)^{-1}
(1+B\sin\theta)\exp\left(-\frac{t}{\tgas}\right)
\end{equation}
where
\begin{equation}
\theta(a)=\omega\ln(a/\ain)-\pi
\end{equation}

Differentiating the expression for $\sigmanogap$, we find that
\begin{equation}
\phi=1-\frac{B\omega\cos\theta}{(1+B\sin\theta)}
\end{equation}

The most extreme (negative) value of $\phi$ occurs when
\begin{equation}
\frac{d\phi}{d\theta}=0
\end{equation}
which occurs when $\sin\theta=-B$.

Thus migration can be halted when
\begin{equation}
\phi\ge\phi_{\rm min}=1-\frac{B\omega}{(1-B^2)^{1/2}}
\end{equation}

Substituting this in Eqn.~\ref{eq-ktrap} we find that the condition for a planet trap to exist is
\begin{equation}
K\le 20\ln\left(\frac{2.05-1.1\phi_{\rm min}}
{3.35-0.1\phi_{\rm min}}\right)
\end{equation}

For the baseline case, we have $B=0.47$ and $\omega=2\pi/\ln 2$. This gives $\phi_{\rm min}=-3.83$ and $K=10.4$. For the planet at $a_p=9$ AU in Figure~5 the disk aspect ratio is $\hgas/a=0.050$. Using this value and $\alpha=10^{-4}$, and the definition of $K$, we find that the maximum planet mass for a trap to exist is $6.0M_\oplus$.

%
%
{}

\end{document}